# (Accepted Manuscript) Reliability modeling and statistical analysis of accelerated degradation process with memory effects and unit-to-unit variability

Shi-Shun Chen [a], Xiao-Yang Li [a, *], Wen-Rui Xie [b]

[a] School of Reliability and Systems Engineering, Beihang University, Beijing 100191, China

[b] School of Mathematics, Jilin University, Changchun 130012, China

Email: css1107@buaa.edu.cn, leexy@buaa.edu.cn, xiewr22@mails.jlu.edu.cn

Corresponding author[*]: Xiao-Yang Li

## Abstract

A reasonable description of the degradation process is essential for credible reliability assessment in accelerated degradation testing. Existing methods usually use Markovian stochastic processes to describe the degradation process. However, degradation processes of some products are non-Markovian due to the interaction with environments. Misinterpretation of the degradation pattern may lead to biased reliability evaluations. Besides, owing to the differences in materials and manufacturing processes, products from the same population exhibit diverse degradation paths, further increasing the difficulty of accurate reliability estimation. To address the above issues, this paper proposes an accelerated degradation model incorporating memory effects and unit-to-unit variability. The memory effect in the degradation process is captured by the fractional Brownian motion, which reflects the non-Markovian characteristic of degradation. The unit-to-unit variability is considered in the acceleration model to describe diverse degradation paths. Then, lifetime and reliability under normal operating conditions are presented. Furthermore, to give an accurate estimation of the memory effect, a new statistical analysis method based on the expectation maximization algorithm is devised. The effectiveness of the proposed method is verified by a simulation case and a real-world tuner reliability analysis case. The simulation case shows that the estimation of the memory effect obtained by the proposed statistical analysis method is much more accurate than the traditional one. Moreover, ignoring unit-to-unit variability can lead to a highly biased estimation of the memory effect and reliability. From the tuner reliability analysis case, the proposed model is superior in both deterministic degradation trend predictions and degradation boundary quantification compared to existing models, which can provide more credible reliability assessment. The code for the simulation case is publicly available at https://github.com/dirge1/FBM_ADT.

## Highlights

- A non-Markovian accelerated degradation model with memory effect is developed.
- Unit-to-unit variability is considered in the accelerated degradation model.
- Ignoring unit-to-unit variability leads to biased estimation of the memory effect.
- A statistical analysis method is proposed via expectation maximization algorithm.
- The proposed statistical analysis method gives a more accurate estimation.

## 1 Introduction

With the continuous progress of design and manufacturing technology, modern products tend to have extremely high reliability and long lifetime. At this point, accelerated degradation testing (ADT) is always employed, in which degradation data are obtained under more severe stress conditions. By modeling accelerated degradation data, the lifetime and reliability of products under normal conditions can be extrapolated and the saving of cost and time can be achieved. In ADT modeling, a reasonable description of the degradation process is vital to support credible lifetime and reliability assessments [1]. Most existing accelerated degradation models hold a general assumption that performance degradation is a memoryless Markovian process with independent increments [2-4]. However, for real engineering products, such as batteries or blast furnace walls [5], degradation typically exhibits non-Markovian properties due to their interaction with environments [6], i.e., the future degradation increment is influenced by the current or historical states. Since traditional Markovian models are inherently memoryless, they fail to describe the degradation of such dynamic systems, leading to an imprecise lifetime and reliability assessments [7]. Hence, building up an accelerated degradation model considering the non-Markovian degradation is essential for enabling accurate extrapolation of lifetime and reliability.

To describe the non-Markovian property of a degradation process, some research introduced a state-dependent function to quantify the influence of the current state on the degradation rate and established different transformed stochastic processes. For instance, Giorgio et al. [8] established the transformed gamma process and derived the conditional distribution of the first passage time. Following this, Giorgio and Pulcini [9] further proposed the transformed Wiener process to describe the non-Markovian degradation process with possibly negative increments. Also, the transformed inverse Gaussian process [10] and the transformed exponential dispersion process [11] were presented and the corresponding parameter estimation methods based on the Bayesian framework were developed. However, for the above transformed stochastic processes, selecting a sensible form of the state-dependent function is challenging in the absence of prior knowledge. Besides, these models only focus on the influence of the current state on the future degradation increment, which may not be sufficient for describing the memory effects of the entire degradation path.

Different from the state-dependent function, fractional Brownian motion (FBM) serves as a proficient mathematical tool for modeling non-Markovian stochastic processes incorporating memory effects [12, 13]. In an FBM, memory effects can be simply quantified by the Hurst exponent. To this end, FBM has drawn much attention in degradation modeling and had widespread applications in remaining useful life (RUL) prediction [14]. Xi et al. [15] first introduced the FBM to capture the memory effects in the degradation process and predicted the RUL of a turbofan engine. Afterward, Xi et al. [16] proposed an improved degradation model by considering both the memory effect and unit-to-unit variability (UtUV). Shao and Si [17] extended the degradation model based on the FBM by considering measurement errors. Xi et al. [18] further developed a multivariate degradation model based on the FBM to describe multivariate stochastic degradation systems with memory effects. The above studies found that compared to the Wiener process-based Markovian degradation models, the FBM-based non-Markovian ones can help for obtaining more accurate RUL predictions when facing degradation with memory effects.

Despite substantial progress in degradation modeling achieved by introducing FBM, their studies ignore the effect of external stresses which are key factors influencing performance degradation. As a result, their modeling is limited to degradation under constant stress levels. In an ADT, degradation data are collected from different stress levels. In order to extrapolate reliability under the normal stress level, the effect of external stresses on performance degradation needs to be quantified accurately. For this purpose, Si et al. [7] developed an ADT model based on the FBM, which incorporated external stresses into degradation modeling. However, there still exists research gaps. In general, the reliability evaluation accuracy of the ADT model depends not only on the precise description of the degradation process, but also on proper uncertainty quantification. In an ADT, we can observe that different items exhibit diverse degradation paths, denoted as UtUV. UtUV stems from variations in materials and manufacturing among different products in the population. Unfortunately, the existing FBM-based accelerated degradation model ignores UtUV. Therefore, their model can only represent the degradation paths of the observed samples, rather than the overall degradation pattern of the population. The author [19] and other scholars [20, 21] have also shown that ignoring UtUV in accelerated degradation modeling results in unreliable reliability assessments in practical applications.

Apart from accelerated degradation modeling, performing statistical analysis on the accelerated degradation data to estimate unknown parameters is also a crucial step. When estimating the unknown parameters in ADT models with UtUV, it is always challenging to get a solution by direct constrained optimization of the overall likelihood function since the source of effect on degradation is diverse (i.e., external stresses, time and uncertain degradation rate). In the literature, a two-step maximum likelihood estimation (MLE) method was generally used to estimate unknown parameters in the Wiener-process based accelerated degradation models with UtUV [22, 23]. However, the two-step MLE method maximizes two partial likelihood functions separately and fails to guarantee the maximization of the overall likelihood function. This exists a risk of deviating from the real law of actual degradation data when dealing with accelerated degradation models with UtUV, especially when the memory effect is incorporated. As demonstrated in the following Section 4.3.1, the estimation of the memory effect obtained by the two-step MLE method is highly biased, which leads to misjudgment of the degradation law and significant deviations in reliability evaluations.

Motivated by the above limitations, a comprehensive degradation model considering memory effects, UtUV and external stresses simultaneously is presented. In addition, a new statistical analysis method is proposed based on the expectation maximization (EM) algorithm [24], where the maximization of the overall likelihood function can be implemented. Compared to existing studies, although Xi et al. [16] has proposed a FBM-based degradation model incorporating UtUV, their model ignored the effect of external stresses on degradation. Therefore, extrapolation of the product reliability under other stress levels cannot be achieved. For the accelerated degradation model proposed in [7], UtUV was ignored in their model, which could result in biased reliability evaluations in practical applications. Moreover, when external stresses and UtUV are both considered in the FBM-based degradation model, the statistical analysis method used in [16] and [7] has difficulty in obtaining a solution, requiring exploration of new statistical analysis methods.

The main contributions of this paper are summarized as follows:

1) An improved FBM-based degradation model is proposed for reliability estimation of accelerated degradation data with memory effects, in which the influence of external stresses are incorporated and the UtUV is quantified.
2) A statistical analysis method is presented based on the EM algorithm, in which the accurate estimation of the memory effect is achieved by maximizing the overall likelihood function rather than maximizing two partial likelihood functions separately.
3) The superiority of the proposed accelerated degradation model and statistical analysis method is verified by comparing existing methods in a numerical case and a real-world tuner reliability analysis case.

The organization of the paper is as follows. Firstly, Section 2 gives the preliminary about the FBM. Then, in Section 3, an ADT model with memory effects and UtUV is proposed, and a statistical analysis method based on the EM algorithm is presented. After that, in Section 4 and Section 5, the practicability and the superiority of the proposed methodology are verified by a simulation case and an engineering case, respectively. Finally, we conclude our work in Section 6.

## 2 Preliminary to fractional Brownian motion

Let a standard FBM denoted by $B_H(t)$, for a given time $t \geq 0$, $B_H(t)$ can be defined as [25]:

$$B_H(t) = \frac{1}{\Gamma\left(H + \frac{1}{2}\right)} \int_{-\infty}^{t} W_H(t-r)\mathrm{d}B(r), \tag{1}$$

where

$$W_H(t-r) = \begin{cases} (t-r)^{H-1/2}, & 0 \leq r \leq t \\ (t-r)^{H-1/2} - (-r)^{H-1/2}, & r < 0; \end{cases} \tag{2}$$

$\Gamma(\cdot)$ is the gamma function formulated by

$$\Gamma(x) = \int_{0}^{\infty} t^{x-1} e^{-t} \mathrm{d}t; \tag{3}$$

$B(\cdot)$ represents the standard Brownian motion; and $H$ is the Hurst exponent with the range of (0, 1). For the standard FBM shown in Eq. (1), the autocorrelation function can be expressed by

$$\mathrm{E}\left[B_H(t)B_H(r)\right] = \frac{1}{2}\left(t^{2H} + r^{2H} - |t-r|^{2H}\right). \tag{4}$$

It can be seen from Eq. (4) that the memory effect is determined by $H$. Depending on the value of $H$, there are three different cases [16]:

1) When $0 < H < 0.5$, the increments of FBM are negatively correlated, which implies that future

increments tend to an opposite tendency of the previous direction and the increment process has short-term memory effect.

2) When $H = 0.5$, the FBM degenerates into the BM, and there is no memory effect in the increment process.

3) When $0.5 < H < 1$, the increments of FBM are positively correlated, which implies that future increments tend to follow the previous direction and the increment process has long-term memory effect.

We can see that the standard BM is a special case of the standard FBM. In addition, for $t \geq 0$, the standard FBM exhibits the following properties:

1) $B_H(0) = 0$;

2) $E\left[B_H(t)\right] = 0$ and $E\left[B_H^2(t)\right] = t^{2H}$;

3) $B_H(t) - B_H(r) \sim B_H(t-r)$ for any $r < t$;

4) $B_H(at) \sim a^H B_H(t)$ for $a > 0$.

# 3 Methodology

## 3.1 Model construction

In order to describe the non-Markovian degradation process with memory effects, a standard FBM shown in Eq. (1) is introduced to construct the ADT model as follows:

$$X(s,t) = f(s,t) + \sigma B_H(t), \tag{5}$$

where $X(\cdot)$ denotes the performance degradation amount; $f(\cdot)$ is a function of stress $s$ and time $t$, which depicts the deterministic performance degradation trend; $\sigma$ represents the diffusion coefficient.

Generally speaking, $f(\cdot)$ in Eq. (5) can be decoupled as a performance degradation rate function $e(s)$ multiplied by a time scale function $\tau(t)$ [19], i.e.:

$$X(s,t) = e(s) \cdot \tau(t) + \sigma B_H(t), \tag{6}$$

where $\tau(t) = t^\beta$, $\beta > 0$ is a common assumption for the timescale function [23]. If $\beta = 1$, then it represents a linear degradation process, otherwise a nonlinear degradation process; $e(s)$ indicates the relationship between degradation rate and external stress (also known as the acceleration model) [26].

In general, $e(s)$ is usually assumed as a log-linear function as follows [27]:

$$e(s) = \exp\left(\alpha_0 + \alpha_1 s^*\right), \tag{7}$$

where $\alpha_0$ and $\alpha_1$ are constant unknown parameters; $s^*$ is the standardized stress. For different type of acceleration models, $s^*$ can be calculated as [28]:

$$s^* = \begin{cases} \dfrac{1/s_0 - 1/s}{1/s_0 - 1/s_H}, & \text{Arrhenius model,} \\[2ex] \dfrac{\ln s - \ln s_0}{\ln s_H - \ln s_0}, & \text{Power law model,} \\[2ex] \dfrac{s - s_0}{s_H - s_0}, & \text{Exponential model,} \end{cases} \tag{8}$$

where $s_0$ denotes the normal stress level; $s_H$ denotes the highest stress levels. The choice of acceleration models mainly depends on the type of accelerated stress. For instance, the Arrhenius model is adopted if the accelerated stress is temperature; if the accelerated stress is humidity, the Power law model is employed [29].

As mentioned in the introduction, UtUV plays a significant role in ADT modeling. To be specific, for items under the same stress level, each item may have its own performance degradation rate due to manufacturing imperfections. As a result, it is reasonable to deduce that each item has unique $e(s)$. Without loss of generality, we make the assumption that the degradation rates of various items adhere to a normal probability distribution. Thus, Eq. (6) can be transformed as:

$$X(s,t) = e(s)\cdot\tau(t) + \sigma B_H(t), e(s) \sim N\left(\mu_e(s), \sigma_e^2(s)\right). \tag{9}$$

Given that $e(s)$ is a random variable, the degradation rate function in Eq. (7) can be reformulated, as indicated by

$$e(s) = \exp\left(\alpha_0 + \alpha_1 s^*\right) = a\exp\left(\alpha_1 s^*\right), a \sim N\left(\mu_a, \sigma_a^2\right), \tag{10}$$

where $a = \exp(\alpha_0)$ is a random variable with a normal probability distribution, $\mu_a > 0$. Then, the distribution of $e(s)$ in Eq. (9) can be derived as:

$$\mu_e(s) = \mu_a \exp\left(\alpha_1 s^*\right),\ \sigma_e(s) = \sigma_a \exp\left(\alpha_1 s^*\right). \tag{11}$$

### 3.2 Lifetime distribution and reliability analysis

The main purpose of ADT is to extrapolate the lifetime and reliability of the product at normal stress $s_0$ by using the degradation data subjected to high stress levels. In accordance with the definition of reliability within belief reliability theory [30], a product is considered to fail when its performance margin less than 0. Since $X$ denotes the performance degradation amount, the performance margin $M$ can be defined as:

$$M(s,t) = X_{th} - X(s,t), \tag{12}$$

where $X_{th}$ denotes the critical threshold of the degradation process. Then, the failure time $T$ of a product can be expressed as:

$$T(s) = \inf\left\{t \middle| M(s,t) < 0\right\}, \tag{13}$$

and the reliability of a product $R$ can be expressed as:

$$R(s,t) = \Pr\left\{T(s) \geq t\right\}, \tag{14}$$

where $\Pr\{\cdot\}$ denotes the probability measure.

For the FBM-based degradation model with UtUV, Xi et al. [16] have developed an approximate analytical lifetime distribution by utilizing weighted random sums. However, the approximate lifetime distribution is only suitable for the degradation process with long-term memory effect, i.e., $H$>0.5. When the degradation process exhibits short-term dependency, the approximate lifetime distribution may lead to biased reliability evaluations. Therefore, we employ the Monte Carlo (MC) method to calculate the lifetime distribution and reliability under specified operating conditions. This approach is not restricted by the type of memory effect. Algorithm 1 gives the detailed procedures of the MC method.

**Algorithm 1**: The MC method to calculate the product lifetime distribution and reliability under specified operating conditions

Step 1. Determine the model parameters $\hat{\boldsymbol{\theta}} = \left[\hat{\mu}_a, \hat{\sigma}_a, \hat{\alpha}_1, \hat{\beta}, \hat{\sigma}, \hat{H}\right]^T$ based on the EM algorithm in the subsequent Section 3.3;

Step 2. According to $\hat{\boldsymbol{\theta}}$, simulate $N$ degradation trajectories under specified stress $s_p$. The $q^{\text{th}}$ degradation trajectory can be obtained as:

2.1 Simulate the $q^{\text{th}}$ standard FBM $B^q_{H=\hat{H}}(t)$ using the FFT algorithm [31] (Detailed steps of the FFT algorithm can be found in [7]);

2.2 Generate a unit-specific $a^q$ according to its distribution $N\left(\hat{\mu}_a, \hat{\sigma}_a^2\right)$;

2.3 Acquire the degradation trajectory as $X_q(s,t) = a^q \exp\left(\hat{\alpha}_1 s_p^*\right) \cdot t^{\hat{\beta}} + \hat{\sigma} B^q_{H=\hat{H}}(t)$.

Step 3. Calculate the product lifetime distribution and reliability under specified stress $s_p$:

3.1 Calculate the failure time of the $N$ simulated degradation trajectories under stress $s_p$ by applying Eq. (13), denoted as $T_q$ $(1 \le q \le N)$;

3.2 Derive the empirical lifetime distribution under stress $s_p$ as $F\left(t \middle| \hat{\boldsymbol{\theta}}, s_p\right) = \frac{1}{N}\sum_{q=1}^{N} 1_t\left(T_q\right)$;

$$1_t\left(T_q\right) = \begin{cases} 1, \text{if } T_q \leqslant t \\ 0, \text{otherwise} \end{cases};$$

3.3 Calculate the product reliability at the specific time $t$ using Eq. (14).

3.3 Statistical analysis

In this paper, we consider a constant stress ADT, which is generally used in practice. The observed ADT data $x_{lij}$ is denoted as the $j^{\text{th}}$ degradation value of unit $i$ subjected to the $l^{\text{th}}$ stress level, $l$ = 1, 2, ..., $k$, $i$ = 1, 2, ..., $n_l$, $j$ = 1, 2, ..., $m_{li}$, where $k$ is the number of stress levels, $n_l$ is the number of test items under the $l^{\text{th}}$ stress level, and $m_{li}$ denotes the number of measurements for unit $i$ subjected to the $l^{\text{th}}$ stress level. The corresponding measurement time is represented by $t_{lij}$. Hereby, the degradation observations of the $i^{\text{th}}$ item tested at $l^{\text{th}}$ stress level can be denoted as $\boldsymbol{x}_{li} = \left[x_{li1}, x_{li2}, \cdots, x_{lim_{li}}\right]^T$. Furthermore, we define $\boldsymbol{\tau}_{li} = \left[\tau\left(t_{li1}\right), \tau\left(t_{li2}\right), \cdots, \tau\left(t_{lim_{li}}\right)\right]^T$ and $\boldsymbol{B}_H^{li} = \left[B_H\left(t_{li1}\right), B_H\left(t_{li2}\right), \cdots, B_H\left(t_{lim_{li}}\right)\right]^T$.

According to Eqs. (9) and (10), the unknown parameters need to be determined are $\boldsymbol{\theta} = \left[\mu_a, \sigma_a, \alpha_1, \beta, \sigma, H\right]^T$. In this section, a statistical analysis approach based on the EM algorithm is presented.

Since each item has a unique degradation rate, for the $i^{\text{th}}$ item tested at $l^{\text{th}}$ stress level, we have

$$\boldsymbol{x}_{li} = a_{li}\boldsymbol{\psi}_{li} + \sigma \boldsymbol{B}_H^{li}, \tag{15}$$

where

$$\boldsymbol{\psi}_{li} = \exp\left(\alpha_1 s_l^*\right)\boldsymbol{\tau}_{li}. \tag{16}$$

The basic idea of the EM algorithm involves replacing unobservable variables with their conditional expectations, where parameter updates at each step can be derived in a closed or simple form [32]. The EM algorithm mainly comprises two steps. Firstly, the expectation of the log-likelihood is calculated with respect to the latent variables, which is called the expectation-step (E-step). Then, the maximizer of this expected likelihood is identified, which is called the maximization step (M-step). These two steps are iteratively repeated until reaching a satisfactory level of convergence.

In Eq. (15), $a_{li}$ is the unobservable variable and needs to be replaced. We denote $\mathbf{\Omega}=\left\{a_{11},\ldots,a_{1n_1},a_{21},\ldots,a_{ln_l}\right\}$. Subsequently, when provided with the complete data including $\boldsymbol{x}$ and $\mathbf{\Omega}$, we can formulate the complete log-likelihood function as:

$$\ln L(\boldsymbol{\theta}|\boldsymbol{x},\mathbf{\Omega})=\sum_{l=1}^{k}\sum_{i=1}^{n_l}\left[\ln p\left(\boldsymbol{x}_{li}|a_{li},\boldsymbol{\theta}\right)+\ln p\left(a_{li}|\boldsymbol{\theta}\right)\right], \tag{17}$$

where $p\left(\boldsymbol{x}_{li}|a_{li},\boldsymbol{\theta}\right)$ denotes the probability density function (PDF) of $\boldsymbol{x}_{li}$ given $a_{li}$ and $\boldsymbol{\theta}$; and $p\left(a_{li}|\boldsymbol{\theta}\right)$ denotes the PDF of $a_{li}$ given $\boldsymbol{\theta}$. According to the ADT models Eqs. (9) and (11), Eq. (17) can be extended as:

$$\begin{aligned}\ln L(\boldsymbol{\theta}|\boldsymbol{x},\mathbf{\Omega})=&-\frac{1}{2}\sum_{l=1}^{k}\sum_{i=1}^{n_l}\left[m_{li}\ln(2\pi)+\ln\left|\boldsymbol{Q}_{li}\right|+\left(\boldsymbol{x}_{li}-a_{li}\boldsymbol{\psi}_{li}\right)^T\boldsymbol{Q}_{li}^{-1}\left(\boldsymbol{x}_{li}-a_{li}\boldsymbol{\psi}_{li}\right)\right]\\&-\frac{1}{2}\sum_{l=1}^{k}\sum_{i=1}^{n_l}\left[\ln(2\pi)+\ln\sigma_a^2+\frac{\left(a_{li}-\mu_a\right)^2}{\sigma_a^2}\right],\end{aligned} \tag{18}$$

where $\boldsymbol{Q}_{li}$ is a $K_{li}\times K_{li}$ dimensional covariance matrix and its $(u, v)^{\text{th}}$ entry can be calculated by

$$\left(\boldsymbol{Q}_{li}\right)_{uv}=\frac{\sigma^2}{2}\left(t_{liu}^{2H}+t_{liv}^{2H}-\left|t_{liu}-t_{liv}\right|^{2H}\right). \tag{19}$$

To facilitate the following statistical analysis, a re-parameterization of the unknown parameter is performed by $\boldsymbol{\Sigma}_{li}=\frac{\boldsymbol{Q}_{li}}{\sigma^2}$. Then, Eq. (18) can be rewritten as:

$$\begin{aligned}\ln L(\boldsymbol{\theta}|\boldsymbol{x},\Omega)=-\frac{1}{2}\sum_{l=1}^{k}\sum_{i=1}^{n_l}\Bigg[&\left(m_{li}+1\right)\ln(2\pi)+\ln\left|\boldsymbol{\Sigma}_{ij}\right|+m_{li}\ln\sigma^2\\&+\frac{\left(\boldsymbol{x}_{li}-a_{li}\boldsymbol{\psi}_{li}\right)^T\boldsymbol{\Sigma}_{li}^{-1}\left(\boldsymbol{x}_{li}-a_{li}\boldsymbol{\psi}_{li}\right)}{\sigma^2}+\ln\sigma_a^2+\frac{\left(a_{li}-\mu_a\right)^2}{\sigma_a^2}\Bigg].\end{aligned} \tag{20}$$

Considering the update of $\boldsymbol{\theta}$ during the iterations in the EM algorithm, let $\boldsymbol{\theta}^{(p)}=\left[\mu_a^{(p)},\sigma_a^{2(p)},\alpha_1^{(p)},\beta^{(p)},\sigma^{2(p)},H^{(p)}\right]^T$ represent the estimations in the $p^{\text{th}}$ step. To calculate the expectation of the log-likelihood function Eq. (20), it is necessary to derive the first and second moments of $a_{li}$ conditional on $\boldsymbol{x}_{li}$ and $\boldsymbol{\theta}^{(p)}$. Notably, it is evident that the distribution of $a_{li}$ conditional on $\boldsymbol{x}_{li}$ and $\boldsymbol{\theta}^{(p)}$ maintains a normal distribution [33]. Within the Bayesian framework, the posterior distribution of $a_{li}$ can be acquired as:

$$
\begin{aligned}
p\left(a_{li}\left|\boldsymbol{x}_{li},\boldsymbol{\theta}^{(p)}\right.\right) &\propto p\left(\boldsymbol{x}_{li}\left|\ a_{li},\boldsymbol{\theta}^{(p)}\right.\right)p\left(a_{li}\left|\boldsymbol{\theta}^{(p)}\right.\right)\\
&\propto \exp\left[-\frac{1}{2}\frac{\left(\boldsymbol{x}_{li}-a_{li}\boldsymbol{\psi}_{li}^{(p)}\right)^{T}\left[\boldsymbol{\Sigma}_{li}^{(p)}\right]^{-1}\left(\boldsymbol{x}_{li}-a_{li}\boldsymbol{\psi}_{li}^{(p)}\right)}{\sigma^{2(p)}}\right]\exp\left[-\frac{\left(a_{li}-\mu_{a}^{(p)}\right)^{2}}{2\sigma_{a}^{2(p)}}\right]\\
&\propto \exp\left\{-\frac{1}{2}\left[\left(\frac{\left[\boldsymbol{\psi}_{li}^{(p)}\right]^{T}\left[\boldsymbol{\Sigma}_{li}^{(p)}\right]^{-1}\boldsymbol{\psi}_{li}^{(p)}}{\sigma^{2(p)}}+\frac{1}{\sigma_{a}^{2(p)}}\right)a_{li}^{2}-2\left(\frac{\boldsymbol{x}_{li}^{T}\left[\boldsymbol{\Sigma}_{li}^{(p)}\right]^{-1}\boldsymbol{\psi}_{li}^{(p)}}{\sigma^{2(p)}}+\frac{\mu_{a}^{(p)}}{\sigma_{a}^{2(p)}}\right)a_{li}\right]\right\}\\
&\propto \exp\left\{-\frac{\left[a_{li}-\left(\boldsymbol{x}_{li}^{T}\left[\boldsymbol{\Sigma}_{li}^{(p)}\right]^{-1}\boldsymbol{\psi}_{li}^{(p)}\sigma_{a}^{2(p)}+\mu_{a}^{(p)}\sigma^{2(p)}\right)/\left(\left[\boldsymbol{\psi}_{li}^{(p)}\right]^{T}\left[\boldsymbol{\Sigma}_{li}^{(p)}\right]^{-1}\boldsymbol{\psi}_{li}^{(p)}\sigma_{a}^{2(p)}+\sigma^{2(p)}\right)\right]^{2}}{2\sigma^{2(p)}\sigma_{a}^{2(p)}/\left(\left[\boldsymbol{\psi}_{li}^{(p)}\right]^{T}\left[\boldsymbol{\Sigma}_{li}^{(p)}\right]^{-1}\boldsymbol{\psi}_{li}^{(p)}\sigma_{a}^{2(p)}+\sigma^{2(p)}\right)}\right\}\\
&\sim N\left(\mu_{li}^{(p)},\sigma_{li}^{2(p)}\right)
\end{aligned}
\tag{21}
$$

with

$$
\mu_{li}^{(p)}=\frac{\boldsymbol{x}_{li}^{T}\left[\boldsymbol{\Sigma}_{li}^{(p)}\right]^{-1}\boldsymbol{\psi}_{li}^{(p)}\sigma_{a}^{2(p)}+\mu_{a}^{(p)}\sigma^{2(p)}}{\left[\boldsymbol{\psi}_{li}^{(p)}\right]^{T}\left[\boldsymbol{\Sigma}_{li}^{(p)}\right]^{-1}\boldsymbol{\psi}_{li}^{(p)}\sigma_{a}^{2(p)}+\sigma^{2(p)}},
\tag{22}
$$

$$
\sigma_{li}^{2(p)}=\frac{\sigma^{2(p)}\sigma_{a}^{2(p)}}{\left[\boldsymbol{\psi}_{li}^{(p)}\right]^{T}\left[\boldsymbol{\Sigma}_{li}^{(p)}\right]^{-1}\boldsymbol{\psi}_{li}^{(p)}\sigma_{a}^{2(p)}+\sigma^{2(p)}}.
\tag{23}
$$

Subsequently, we employ the EM algorithm to estimate $\boldsymbol{\theta}$ iteratively. Firstly, the expectation of $\ln L(\boldsymbol{\theta}|\boldsymbol{x},\Omega)$ with regard to $\boldsymbol{\Omega}$ is computed in the E-step. According to Eqs. (20), (22) and (23), we can get

$$
\begin{aligned}
Q\left(\boldsymbol{\theta}\left|\boldsymbol{x},\boldsymbol{\theta}^{(p)}\right.\right) &= E_{\boldsymbol{\Omega}|\boldsymbol{\theta}^{(p)}}[\ln L(\boldsymbol{\theta}|\boldsymbol{x},\boldsymbol{\Omega})]\\
&=-\frac{1}{2}\sum_{l=1}^{k}\sum_{i=1}^{n_l}\left\{\left(m_{li}+1\right)\ln(2\pi)+\ln\left|\boldsymbol{\Sigma}_{li}\right|+m_{li}\ln\sigma^{2}\right.\\
&\quad+\frac{\boldsymbol{x}_{li}^{T}\boldsymbol{\Sigma}_{li}^{-1}\boldsymbol{x}_{li}-2\mu_{li}^{(p)}\boldsymbol{x}_{li}^{T}\boldsymbol{\Sigma}_{li}^{-1}\boldsymbol{\psi}_{li}+\left(\left[\mu_{li}^{(p)}\right]^{2}+\sigma_{li}^{2(p)}\right)\boldsymbol{\psi}_{li}^{T}\boldsymbol{\Sigma}_{li}^{-1}\boldsymbol{\psi}_{li}}{\sigma^{2}}\\
&\quad\left.+\frac{\left[\mu_{li}^{(p)}\right]^{2}+\sigma_{li}^{2(p)}-2\mu_{li}^{(p)}\mu_{a}+\mu_{a}^{2}}{\sigma_{a}^{2}}+\ln\sigma_{a}^{2}\right\}.
\end{aligned}
\tag{24}
$$

In the M-step, the first partial derivatives of $Q\left(\boldsymbol{\theta}\left|\boldsymbol{x},\boldsymbol{\theta}^{(p)}\right.\right)$ with respect to $\mu_a$, $\sigma_a^2$ and $\sigma^2$ can be derived, respectively. Then, by equating each derivative to zero, we can obtain

$$\mu_a^{(p+1)} = \frac{1}{\sum_{l=1}^{k} n_l} \sum_{l=1}^{k} \sum_{i=1}^{n_l} \mu_{li}^{(p)}, \tag{25}$$

$$\sigma_a^{2(p+1)} = \frac{\sum_{l=1}^{k} \sum_{i=1}^{n_l} \left( \left[ \mu_{li}^{(p)} \right]^2 + \sigma_{li}^{2(p)} - 2\mu_{li}^{(p)} \mu_a^{(p+1)} + \left[ \mu_a^{(p+1)} \right]^2 \right)}{\sum_{l=1}^{k} n_l}, \tag{26}$$

$$\sigma^{2(p+1)} = \frac{\sum_{l=1}^{k} \sum_{i=1}^{n_l} \left( \boldsymbol{x}_{li}^T \boldsymbol{\Sigma}_{li}^{-1} \boldsymbol{x}_{li} - 2\mu_{li}^{(p)} \boldsymbol{x}_{li}^T \boldsymbol{\Sigma}_{li}^{-1} \boldsymbol{\psi}_{li} + \left( \left[ \mu_{li}^{(p)} \right]^2 + \sigma_{li}^{2(p)} \right) \boldsymbol{\psi}_{li}^T \boldsymbol{\Sigma}_{li}^{-1} \boldsymbol{\psi}_{li} \right)}{\sum_{l=1}^{k} \sum_{i=1}^{n_l} m_{li}}. \tag{27}$$

Note that $\mu_a^{(p+1)}$ and $\sigma_a^{2(p+1)}$ can be directly determined based on Eqs. (22) and (23), while $\sigma^{2(p+1)}$ depends on the values of $\alpha_1^{(p+1)}$, $\beta^{(p+1)}$ and $H^{(p+1)}$. Upon substitution of $\mu_a^{(p+1)}$, $\sigma_a^{2(p+1)}$ and $\sigma^{2(p+1)}$ into Eq. (24), we get the profile likelihood function as:

$$\begin{aligned} & Q\left( \alpha_1, \beta, H \middle| \boldsymbol{x}, \mu_a^{(p+1)}, \sigma_a^{2(p+1)} \right) \\ & = -\frac{1}{2} \sum_{l=1}^{k} \sum_{i=1}^{n_l} \left\{ \left( m_{li} + 1 \right) \left[ \ln(2\pi) + 1 \right] + \ln \left| \boldsymbol{\Sigma}_{li} \right| + \ln \sigma_a^{2(p+1)} + m_{li} \ln \sigma^{2(p+1)} \right\}. \end{aligned} \tag{28}$$

Then, by maximizing the profile likelihood function, the values of $\alpha_1^{(p+1)}$, $\beta^{(p+1)}$ and $H^{(p+1)}$ can be obtained. Subsequently, the value of $\sigma^{2(p+1)}$ can be determined by substituting $\alpha_1^{(p+1)}$, $\beta^{(p+1)}$ and $H^{(p+1)}$ into Eq. (27). The iterations are typically stopped when the relative change in parameter estimations drops below a predefined threshold. The complete algorithm is outlined in Algorithm 2.

| **Algorithm 2**: Parameter Estimations Using the EM Algorithm |
| --- |
| Step 1. Initialize $\boldsymbol{\theta}^{(0)} = \left[ \mu_a^{(0)}, \sigma_a^{(0)}, \alpha_1^{(0)}, \beta^{(0)}, \sigma^{(0)}, H^{(0)} \right]^T$ and $\varepsilon$. Set $p = 0$.<br>Step 2. Get $\boldsymbol{\theta}^{(p+1)}$ by using the EM algorithm<br>**E-step:**<br>2.1 Calculate $\mu_{li}^{(p)}$ and $\sigma_{li}^{2(p)}$ by using Eqs. (22) and (23);<br>**M-step:**<br>2.2 Update $\mu_a^{(p+1)}$ and $\sigma_a^{2(p+1)}$ by using Eqs. (25) and (26);<br>2.3 Update $\alpha_1^{(p+1)}$, $\beta^{(p+1)}$ and $H^{(p+1)}$ by maximizing the profile likelihood function Eq. (28) through a three-dimensional search;<br>2.4 Update $\sigma^{2(p+1)}$ by substituting $\alpha_1^{(p+1)}$, $\beta^{(p+1)}$ and $H^{(p+1)}$ into Eq. (27). |

| Step 3. If $\left|\boldsymbol{\theta}^{(p+1)}-\boldsymbol{\theta}^{(p)}\right| \le \varepsilon$, obtain $\hat{\boldsymbol{\theta}}=\left[\mu_a^{(p+1)},\sigma_a^{2(p+1)},\alpha_1^{(p+1)},\beta^{(p+1)},\sigma^{2(p+1)},H^{(p+1)}\right]^T$. Otherwise, go to Step 2. |
|---|

In addition to the point estimation of $\boldsymbol{\theta}$ derived from Algorithm 2, it is equally important to consider the interval estimation of $\boldsymbol{\theta}$. Interval estimation provides a range of plausible values within which the true parameter is likely to lie, offering a complete understanding of the estimation uncertainty. While the asymptotic normality of maximum likelihood function is commonly used for interval estimation, deriving the expected Fisher information matrix of the proposed model is challenging due to the existence of memory effects, stress acceleration, and UtUV. Thus, non-parametric bootstrap method is employed to acquire the interval estimation. Since the degradation processes of different items are independent, a resampling procedure with replacement is carried out on the degradation processes of items under each stress condition. Then, $\hat{\boldsymbol{\theta}}$ of the resampled data is obtained based on the EM algorithm. After repeating this procedure $M$ times, we get $M$ bootstrap estimates for different resampled data, denoted as $\left\{\hat{\boldsymbol{\theta}}_1^*,\ldots,\hat{\boldsymbol{\theta}}_M^*\right\}$. Then, the interval estimation at a given confidence level (CL) for each parameter can be derived by calculating the percentiles from $\left\{\hat{\boldsymbol{\theta}}_1^*,\ldots,\hat{\boldsymbol{\theta}}_M^*\right\}$. Besides, the interval estimation of reliability can be derived from $\left\{\hat{\boldsymbol{\theta}}_1^*,\ldots,\hat{\boldsymbol{\theta}}_M^*\right\}$. The complete procedure for interval estimation is outlined in Algorithm 3.

| **Algorithm 3**: Interval Estimation of Parameters and Reliability Using the Non-parametric Bootstrap |
|---|
| Step 1. Obtain $M$ bootstrap estimates of $\boldsymbol{\theta}$ based on the original ADT data. The $q^{\text{th}}$ estimate $\hat{\boldsymbol{\theta}}_q^*$ can be obtained as:<br>1.1 Get the resampled degradation data across all the stress levels. For the $l^{\text{th}}$ stress level, randomly sample $n_l$ items with replacement from the original items under the $l^{\text{th}}$ stress level.<br>1.2 Acquire the estimate $\hat{\boldsymbol{\theta}}_q^*$ of the resampled data based on Algorithm 2.<br>Step 2. Obtain the interval estimation of each parameter in $\boldsymbol{\theta}$ at a specific $(1-\alpha)$ CL. For the $p^{\text{th}}$ parameter in $\boldsymbol{\theta}$, which is denoted as $\theta^p$, its interval estimation can be derived as:<br>2.1 Sort all the estimates of $\theta^p$ from smallest to largest based on $\left\{\hat{\theta}_1^{p*},\ldots,\hat{\theta}_M^{p*}\right\}$.<br>2.2 The interval estimation of $\theta^p$ at $(1-\alpha)$ CL is given by $\left(\theta_{\alpha/2}^p,\theta_{1-\alpha/2}^p\right)$, where $\theta_{\alpha/2}^p$ and $\theta_{1-\alpha/2}^p$ are the $(100\alpha/2)^{\text{th}}$ and $[100(1-\alpha/2)]^{\text{th}}$ percentiles of the sorted parameter estimate, respectively.<br>Step 3. Calculate the interval estimation of reliability at a specific $(1-\alpha)$ CL:<br>3.1 Obtain $M$ reliability estimations according to Algorithm 1 based on $M$ bootstrap estimates of $\boldsymbol{\theta}$, denoted as $\left\{R_1,\ldots,R_M\right\}$. |

3.2 Sort the reliability estimation from smallest to largest based on $\{R_1,...,R_M\}$.

3.3 The interval estimation of reliability at (1−$\alpha$) CL is given by $(R_{\alpha/2}, R_{1-\alpha/2})$, where $R_{\alpha/2}$ and $R_{1-\alpha/2}$ are the $(100\alpha/2)^{th}$ and $[100(1-\alpha/2)]^{th}$ percentiles of the sorted reliability estimation, respectively.

# 4 Simulation study

In this section, the proposed method is illustrated through a numerical simulation case. Furthermore, the superiority of the proposed parameter estimation method and the significance of the proposed ADT model are demonstrated.

## 4.1 Simulation settings

Detailed information on the constant-stress ADT simulation case is shown in Table 1. The degradation process exhibits short-term dependency because $H = 0.1$. Specifically, the sample sizes for each stress level ($N$) are set as 6 and the measurements for each sample ($M$) are set as 10.

Table 1 Information for simulation configuration.

| Content | Values |
|---|---|
| Accelerated stress level (Temperature/°C) | 80, 100, 120 |
| Normal stress level (°C) | 40 |
| Degradation model | $X(s,t) = a\exp\left(\alpha_1 s^*\right)\cdot t^\beta + \sigma B_H(t), a \sim N\left(\mu_a, \sigma_a^2\right).$ |
| Acceleration model | Arrhenius model |
| Parameter value: $\mu_a, \sigma_a, \alpha_1, \beta, \sigma, H$ | 1e-5, 2e-6, 2.5, 1.5, 0.1, 0.1 |
| Inspection interval (h) | 100 |
| Failure threshold | 5 |

## 4.2 Results and analysis

In this case, the initial value of $\boldsymbol{\theta}$ is taken from the estimations of the two-step MLE method. The implementation details of the two-step MLE method are given in the Appendix. The terminated threshold $\varepsilon$ for the EM algorithm iteration is set as 0.01, and the number of bootstrap estimates $M$ is set as 1000. Then, according to Algorithm 2 and Algorithm 3, we obtain the point and interval estimations of unknown parameters at 90% CL as shown in Table 2. It can be seen that the point estimations are basically close to the true values and the true values lie within the region of interval, which shows the effectiveness of the proposed statistical analysis method.

Table 2 Point and interval estimation results of $\boldsymbol{\theta}$ in the simulation case.

| Parameters | Point estimations | 90% confidence interval estimations |
|---|---|---|
| $\mu_a$ | 1.293e-5 | (7.761e-6, 2.049e-5) |
| $\sigma_a$ | 2.269e-6 | (1.007e-6, 3.826e-6) |

| $\alpha_1$ | 2.700 | (2.333, 3.052) |
|---|---|---|
| $\beta$ | 1.452 | (1.398, 1.513) |
| $\sigma$ | 0.081 | (0.058, 0.133) |
| $H$ | 0.117 | (0.012, 0.181) |

Furthermore, to show the effectiveness of the ADT model, the predicted deterministic degradation trends based on the point estimations, along with the upper and lower boundaries at 90% CL derived from 10000 simulated degradation paths are compared with the observed data. As depicted in Fig. 1, the predicted deterministic degradation trends primarily lie in the observed data across all accelerated stress levels, and the boundaries nicely envelope the observations. This indicates that the proposed ADT model is capable of describing the degradation law under different stress levels.

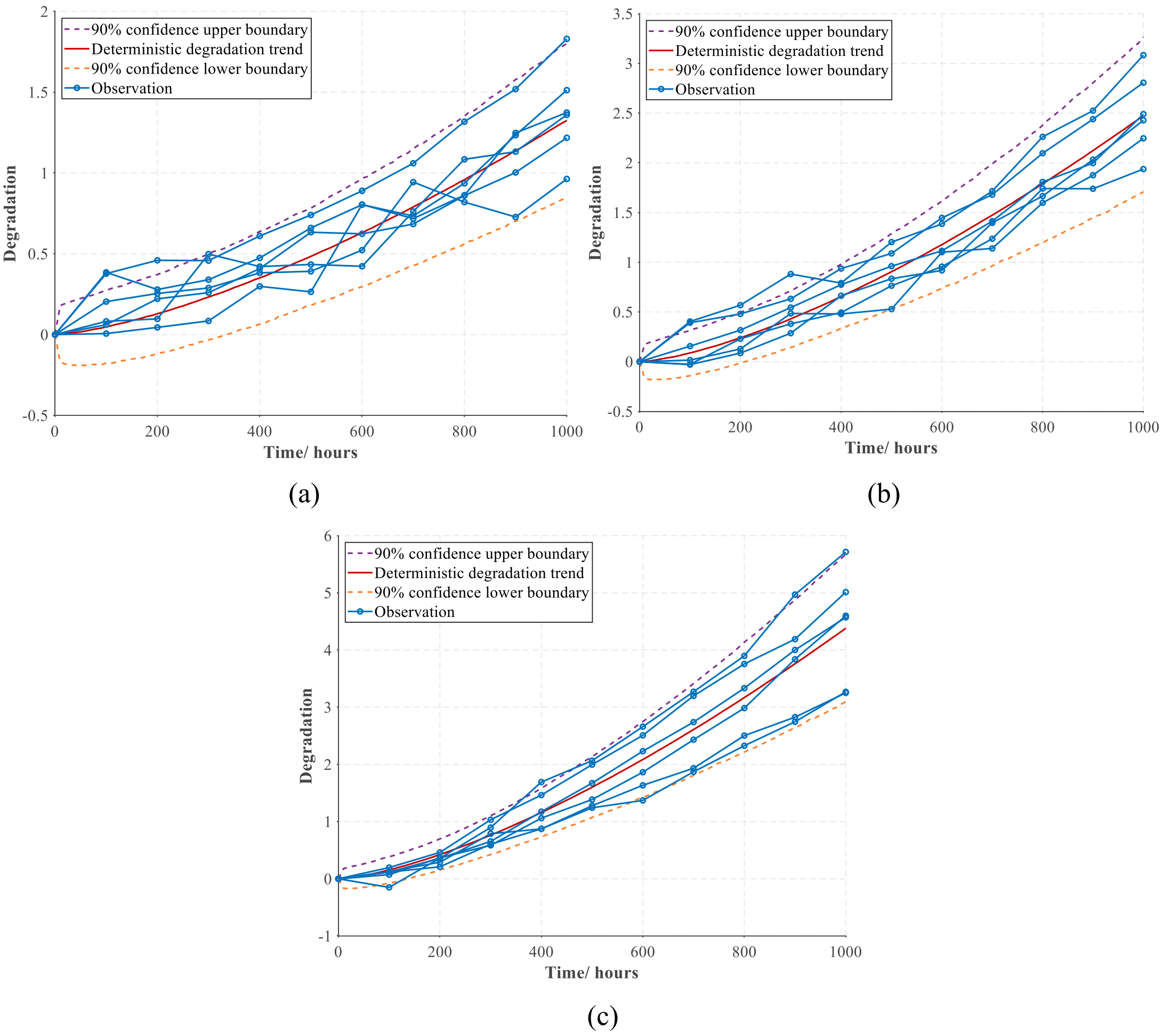

Fig. 1 Deterministic degradation trend, boundaries and observations: (a) 80°C (b) 100°C (c) 120°C.

Next, when the critical threshold $X_{th}$ is 5, we can calculate the point and interval estimation of reliability under the normal use condition (40°C) according to Algorithm 1 and Algorithm 3, as shown in Fig. 2. From Fig. 2, the reliability of the product for a given operating time can be obtained. For example, when the product is operated for 4200 hours, the lower boundary of reliability at 90% CL is 0.99. This result can provide guidance for the development of maintenance and warranty strategies.

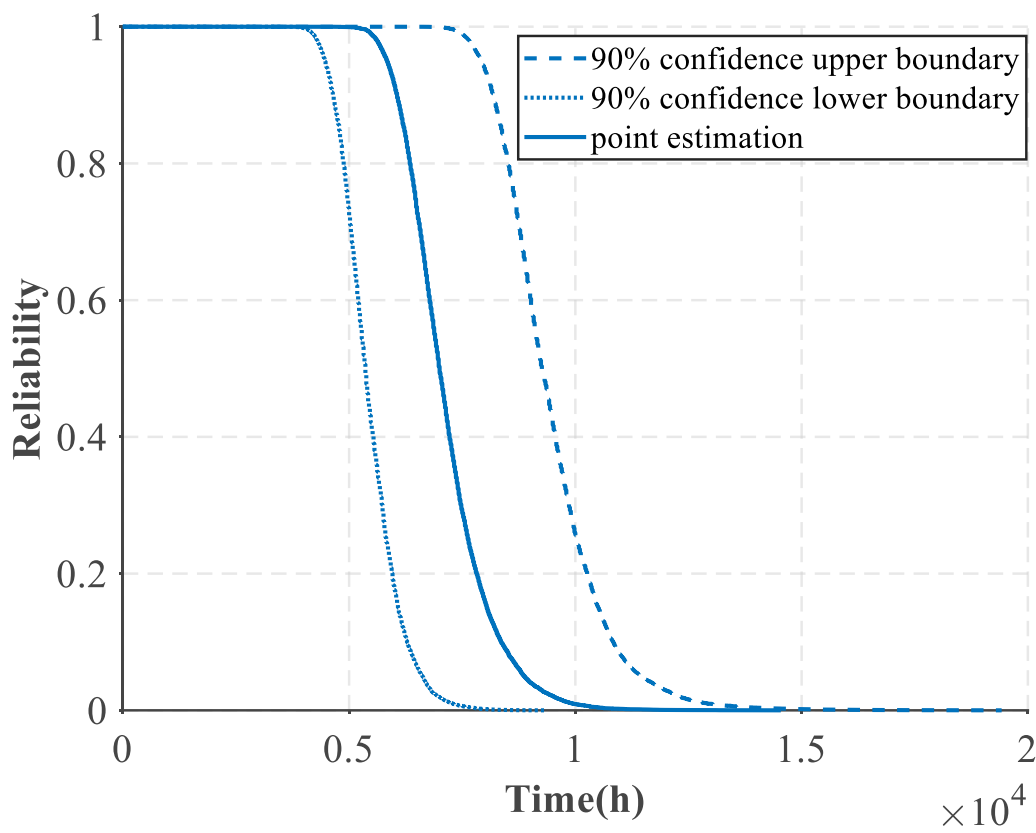

Fig. 2 Reliability of the product under normal stress level (40°C).

**Remark 1**: It is worth mentioning that the sources of uncertainty for the degradation and reliability confidence intervals in Figs. 1 and 2 are different. For the degradation prediction in Fig. 1, the uncertainty is originated from FBM and UtUV in the ADT model, which describes the inherent uncertainty of the degradation process for a batch of products. This type of uncertainty is objective and cannot be eliminated. On the other hand, the uncertainty of reliability estimation in Fig. 2 is originated from the uncertainty in parameter estimations. This type of uncertainty is mainly caused by insufficient sample size. In general, increasing sample size will narrow the range of confidence intervals. In Fig. 1, the goal of quantifying the inherent uncertainty for a batch of products is to compare it with the dispersion of the real data, thereby validating the effectiveness of the ADT model in uncertainty quantification. Whereas in Fig. 2, quantifying the uncertainty of the parameter estimates is chosen in order to prevent the risk of decision making due to estimation bias.

### 4.3 Discussions

#### 4.3.1 Comparison of statistical analysis methods

In order to illustrate the superiority of the proposed statistical analysis method compared to the existing two-step MLE method, in this subsection, nine combinations of $(N,M)$ are chosen to generate simulated ADT data. Other simulation settings are the same as those in Table 1. Considering the randomness of data generation, we simulated ADT data 1000 times for each combination and obtain the mean values of parameter estimations. Then, the relative error (RE) between the estimations and the true values is adopted as a metric to compare the performance of both methods. The RE of the parameter estimation is given by

$$RE = \sum \left| \frac{\hat{\theta} - \theta}{\theta} \right|. \tag{29}$$

Table 3 gives the mean values of unknown parameter estimations obtained by two methods under nine combinations of sample sizes and measurements, and Fig. 3 illustrates the RE for two methods under different combinations. From Table 3 and Fig. 3, we can get the following results:

- For the two-step MLE method, when the number of measurements is limited, the estimation bias of $H$ is significantly high, reaching nearly 100%. As the number of measurements increases, the estimate of $H$ approaches the true value but remains unsatisfactory, with a bias around 40%. On the other hand, for the proposed statistical analysis method based on the EM algorithm, even with a limited number of measurements, the estimated value of $H$ remains very close to the true value, with a bias of less than 5%. This demonstrates the crucial role of maximizing the overall likelihood function for estimating the memory effect accurately, particularly when measurements are scarce.

- With the increase in sample size and the number of measurements, the *RE* obtained by the EM algorithm decreases, demonstrating that increasing information in ADT experiments can enhance the precision of parameter estimation effectively.
- In all combinations, the *RE* obtained by the EM algorithm is much better than those of the two-step MLE method, proving its superiority in estimating the parameters of the ADT model with memory effects and UtUV.

Table 3 Mean value of parameter estimations for two methods under nine combinations of sample sizes and measurements.

| $(N,M)$ | Method | $\mu_a$ | $\sigma_a$ | $\alpha_1$ | $\beta$ | $\sigma$ | $H$ | *RE* |
|---|---|---|---|---|---|---|---|---|
| (6,10) | S1 | 1.100e-5 | 2.221e-6 | 2.469 | 1.5 | 0.157 | 0.001 | 1.779 |
| | S2 | 1.102e-5 | 1.998e-6 | 2.483 | 1.498 | 0.106 | 0.095 | **0.235** |
| (6,20) | S1 | 1.041e-5 | 1.938e-6 | 2.493 | 1.5 | 0.144 | 0.03 | 1.209 |
| | S2 | 1.041e-5 | 1.918e-6 | 2.493 | 1.5 | 0.103 | 0.095 | **0.168** |
| (6,30) | S1 | 1.027e-5 | 2.553e-6 | 2.493 | 1.5 | 0.124 | 0.059 | 0.959 |
| | S2 | 1.027e-5 | 1.925e-6 | 2.493 | 1.5 | 0.102 | 0.097 | **0.116** |
| (12,10) | S1 | 1.068e-5 | 2.257e-6 | 2.477 | 1.499 | 0.158 | 4.484e-5 | 1.790 |
| | S2 | 1.070e-5 | 2.049e-6 | 2.489 | 1.497 | 0.105 | 0.095 | **0.134** |
| (12,20) | S1 | 1.016e-5 | 1.848e-6 | 2.497 | 1.5 | 0.142 | 0.032 | 1.192 |
| | S2 | 1.016e-5 | 1.959e-6 | 2.497 | 1.5 | 0.102 | 0.098 | **0.074** |
| (12,30) | S1 | 1.016e-5 | 2.288e-6 | 2.494 | 1.5 | 0.122 | 0.061 | 0.772 |
| | S2 | 1.016e-5 | 1.968e-6 | 2.494 | 1.5 | 0.101 | 0.099 | **0.055** |
| (18,10) | S1 | 1.064e-5 | 2.270e-6 | 2.466 | 1.499 | 0.159 | 1.531e-8 | 1.802 |
| | S2 | 1.067e-5 | 2.066e-6 | 2.481 | 1.497 | 0.102 | 0.098 | **0.151** |
| (18,20) | S1 | 1.013e-5 | 2.007e-6 | 2.497 | 1.5 | 0.141 | 0.033 | 1.101 |
| | S2 | 1.013e-5 | 1.981e-6 | 2.497 | 1.5 | 0.101 | 0.099 | **0.049** |
| (18,30) | S1 | 1.003e-5 | 1.968e-6 | 2.506 | 1.5 | 0.122 | 0.061 | 0.624 |
| | S2 | 1.003e-5 | 1.960e-6 | 2.506 | 1.5 | 0.101 | 0.099 | **0.044** |

*Note*: S1 represents the two-step MLE method, and S2 represents the proposed statistical analysis method based on the EM algorithm.

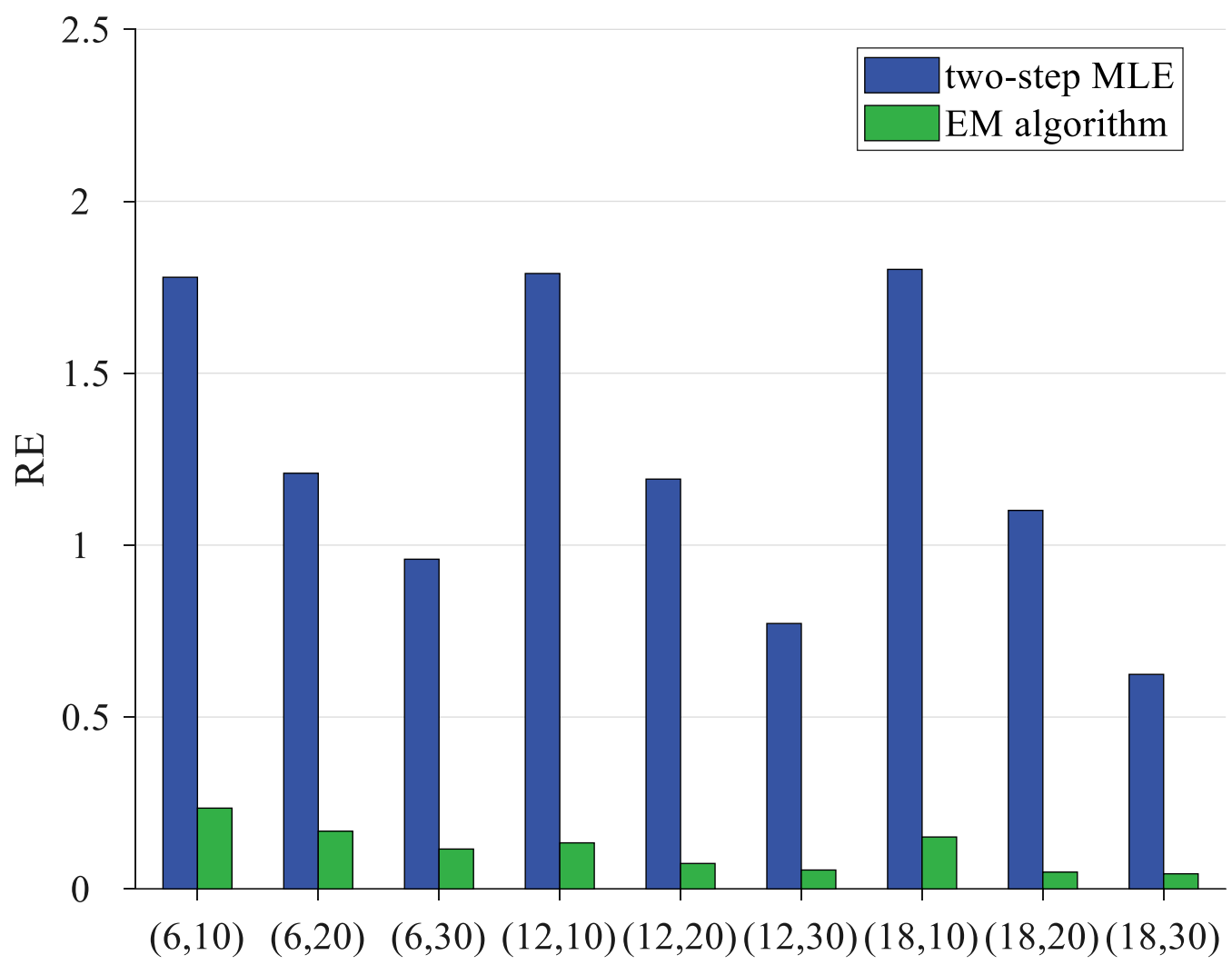

Fig. 3 RE for parameter estimations of two statistical analysis methods under nine combinations of sample sizes and measurements.

#### 4.3.2 Comparison of ADT models

The proposed ADT model is described by Eqs. (8), (9), and (11), which is denoted as $M_0$. It should be noted that the proposed ADT model has three special cases as follows:

- When the UtUV is not considered, i.e., $a$ is a constant, it degenerates into the model in [7], which is denoted as $M_1$.

$$M_1 : X(s,t) = a\exp\left(\alpha_1 s^*\right)t^\beta + \sigma B_H(t). \tag{30}$$

- When the memory effects in the degradation process are not considered, i.e., $H=0.5$, it degenerates into the model in [22], which is denoted as $M_2$.

$$M_2 : X(s,t) = a\exp\left(\alpha_1 s^*\right)t^\beta + \sigma B(t), a \sim N\left(\mu_a, \sigma_a^2\right). \tag{31}$$

- When the UtUV and the memory effects are both not considered, i.e., $a$ is a constant and $H=0.5$, it degenerates into the model in [34], which is denoted as $M_3$.

$$M_3 : X(s,t) = a\exp\left(\alpha_1 s^*\right)t^\beta + \sigma B(t). \tag{32}$$

In order to illustrate the superiority of the proposed ADT model compared to existing ADT models, in this subsection, we set the combination of ($N$,$M$) as (12,30) to generate ADT simulation data. Other simulation settings are the same as those in Table 1. Considering the randomness of data generation, we simulated ADT data 1000 times for each combination and obtain the mean values of parameter estimations. For the model $M_0$ and $M_2$, the unknown parameters are estimated based on the EM algorithm. For the model $M_1$ and $M_3$, the parameters are obtained by directly maximizing the log-likelihood function. Then, we calculate the maximum value of the log-likelihood function $l_{\max}$ and the Akaike information criterion (AIC) for each ADT model. The AIC is expressed as [35]

$$AIC = -2l_{\max} + 2n_p, \tag{33}$$

where $n_p$ is the number of unknown parameters. The smaller the AIC, the better the model fits.

Table 4 gives the mean values of unknown parameter estimations for the four ADT models. From $l_{\max}$ and AIC values, $M_0$ is the most effective model for describing the ADT data, followed by model $M_2$ and $M_1$, and model $M_3$ is the worst. Analyzing the reasons, the model $M_2$ and $M_1$ partially consider the UtUV and the memory effects, respectively. Therefore, they both perform worse than the model $M_0$, which considers the UtUV and the memory effects simultaneously. Moreover, model $M_1$ even gets wrong degradation law, i.e., the degradation process exhibits long-term dependency, which is due to the lack of consideration of UtUV. As for the model $M_3$, both the UtUV and the memory effects are not considered, which leads to the poorest fitting results.

Table 4  Mean value of parameter estimations for the four ADT models.

| Model | $\mu_a$ | $\sigma_a$ | $\alpha_1$ | $\beta$ | $\sigma$ | $H$ | $l_{\max}$ | AIC |
|---|---|---|---|---|---|---|---|---|
| $M_0$ | 1.003e-5 | 1.960e-6 | 2.506 | 1.5 | 0.101 | 0.099 | **811.827** | **-1611.65** |
| $M_1$ | 9.018e-6 | / | 2.522 | 1.512 | 0.013 | 0.583 | 465.075 | -920.15 |
| $M_2$ | 1.032e-5 | 1.851e-6 | 2.501 | 1.497 | 0.016 | / | 605.479 | -1200.96 |
| $M_3$ | 8.598e-6 | / | 2.534 | 1.522 | 0.019 | / | 430.897 | -853.79 |

In addition, the product reliability under normal stress level (40°C) obtained by the four ADT models are compared with the real values as illustrated in Fig. 4. From Fig. 4, the model $M_0$ gives an accurate reliability assessment while the other three ADT models show significant bias, proving the significance of the proposed ADT model.

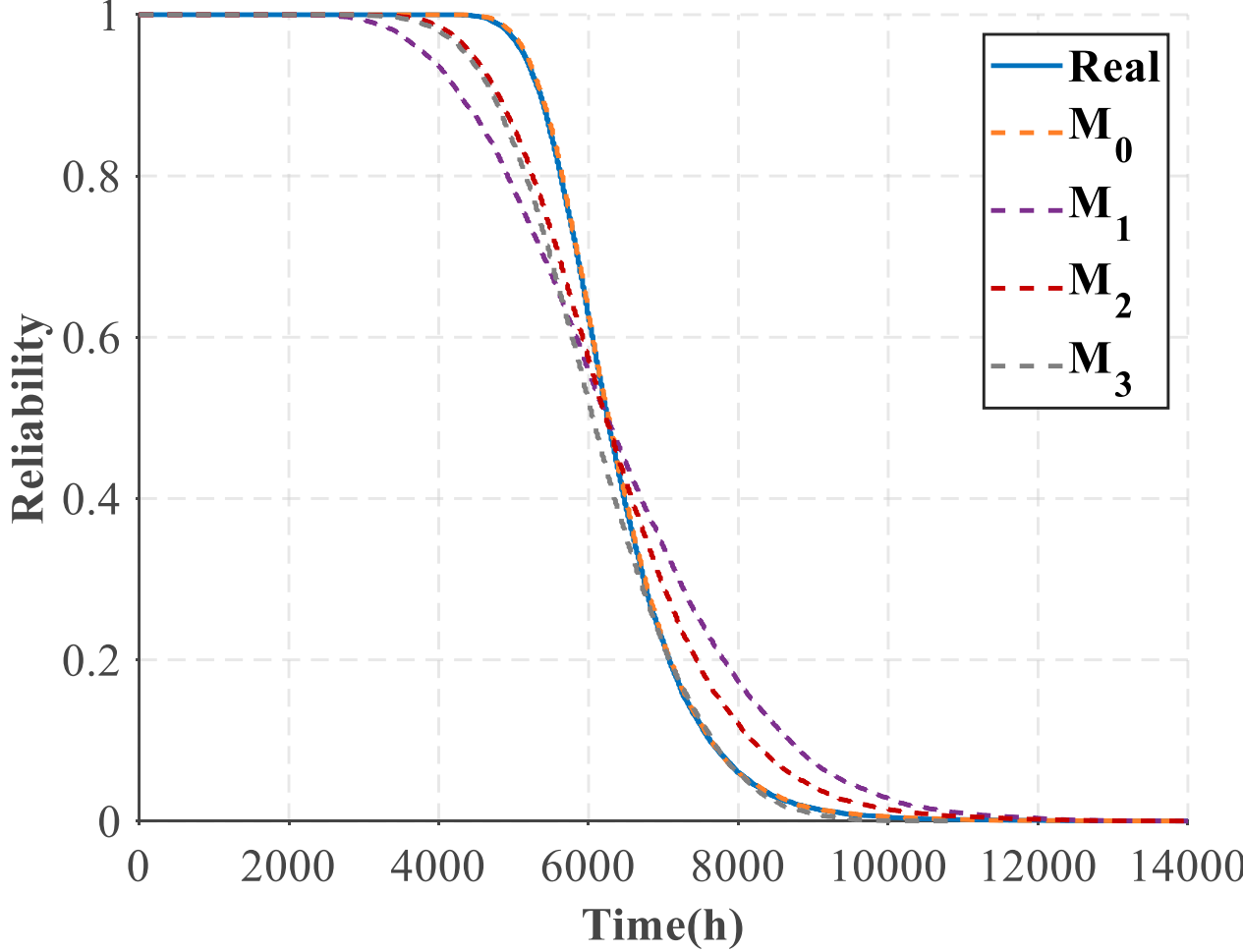

Fig. 4  Reliability of the product under normal stress level (40°C) for the four ADT models with the real values.

## 5 Engineering application

In this section, we take the accelerated degradation data of a tuner to show the superiority of the proposed ADT model compared to other ADT models.

### 5.1 Data description

A tuner is a long-lifespan microwave electronic assembly designed primarily for reception, filtering, amplification and gain control of cable signals. Previous investigations into tuner failures have shown the significant impact of temperature on the degradation process [36]. To evaluate the reliability of the tuner, a constant-stress ADT based on temperature was performed under 4 stress levels. And, the key performance parameter, noise, was measured every ten hours by a computerized measuring system. Fig. 5 illustrates detailed degradation data, which is sourced from Fig. 6 in [36].

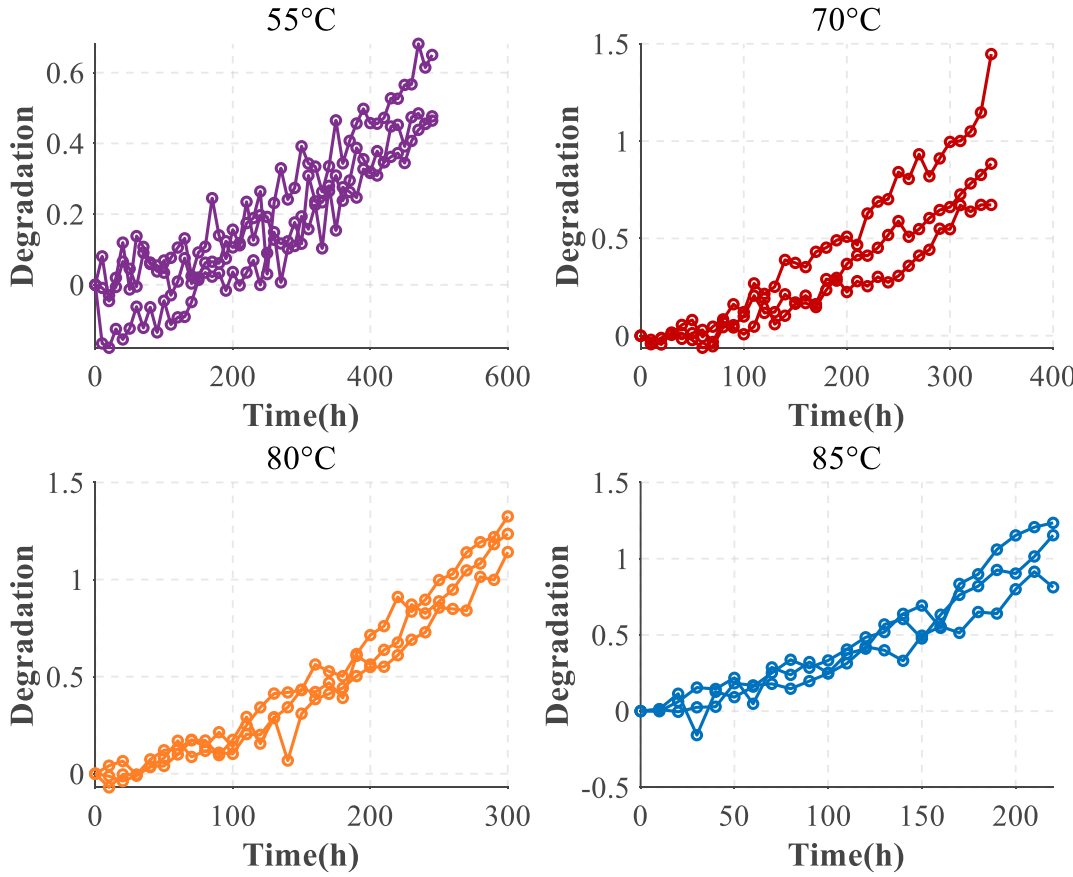

Fig. 5 Accelerated degradation data of a tuner under 4 stress levels.

## 5.2 Modeling and analysis

The Arrhenius model is adopted to build up the acceleration model since the applied accelerated stress is temperature. Then, stress normalization is employed according to Eq. (8), where the normal stress level is 20°C. Next, the initial values of the unknown parameters are set based on the result of the two-step MLE method, and the terminated threshold $\varepsilon$ for the EM algorithm iteration is set as 0.01. Subsequently, according to Algorithm 2, we obtain the estimations of unknown parameters as shown in Table 5.

Table 5 Point and interval estimation results of $\boldsymbol{\theta}$ in the tuner case.

| Parameters | Point estimations | 90% confidence interval estimations |
|---|---|---|
| $\mu_a$ | 1.073e-6 | (6.623e-7, 1.856e-6) |
| $\sigma_a$ | 1.894e-7 | (8.463e-8, 3.399e-7) |
| $\alpha_1$ | 4.667 | (4.304, 5.021) |
| $\beta$ | 1.691 | (1.642, 1.725) |
| $\sigma$ | 0.073 | (0.069, 0.077) |
| $H$ | 2.497e-8 | (1.307e-9, 0.009) |

From Table 5, it can be deduced that the degradation process of the tuner exhibits short-term dependency because $H < 0.5$. The short-term memory effect in the degradation process is also reported in [37]. Besides, it can be found that $H$ is close to zero in our case. At this point, the FBM is close to a logarithmically correlated Gaussian random process [38], which has been studied and employed in financial mathematics [39].

Then, we generate 10000 degradation paths according to the point estimations in Table 5. The predicted deterministic degradation trends and boundaries of performance with temperature and time are plotted in Fig. 6 (a). When the critical threshold $X_{th}$ is 7, the performance margin can be calculated according to Eq. (12). Fig. 6 (b) illustrates the predicted deterministic degradation trends and boundaries of margin with temperature and time. It can be seen from the above figures that the degradation and corresponding boundary width increase with the increase of time and temperature.

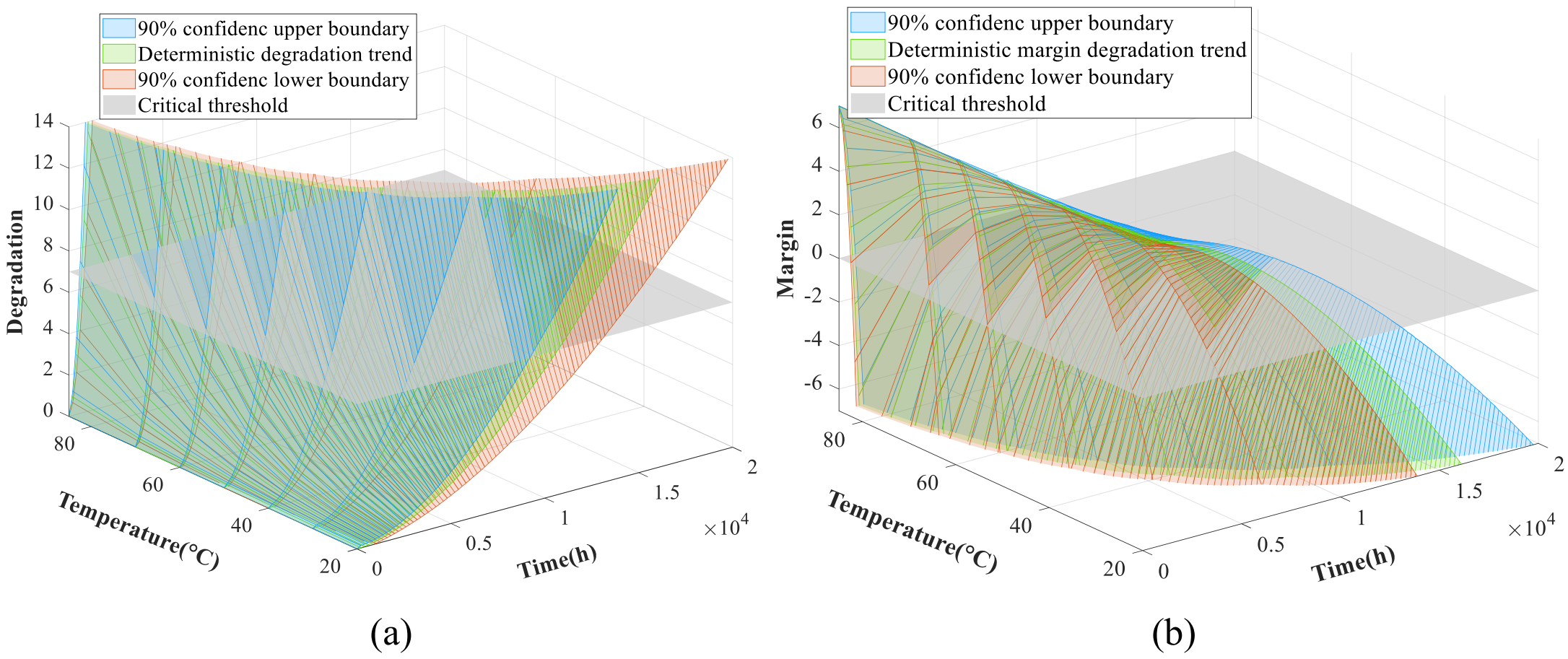

Fig. 6 Deterministic degradation trends and boundaries of tuner performance and margin with temperature and time.

Next, according to Algorithm 1, we can deduce the reliability under different temperature and time as shown in Fig. 7 (a). Specially, the reliability of the tuner under the normal use condition (20°C) with 90% confidence interval is illustrated in Fig. 7 (b). From Fig. 7 (b), we can get some valuable information for the development of maintenance and warranty strategies. For example, when the tuner is operated for 6000 hours, the lower boundary of reliability of the tuner at 90% CL is nearly 0.99.

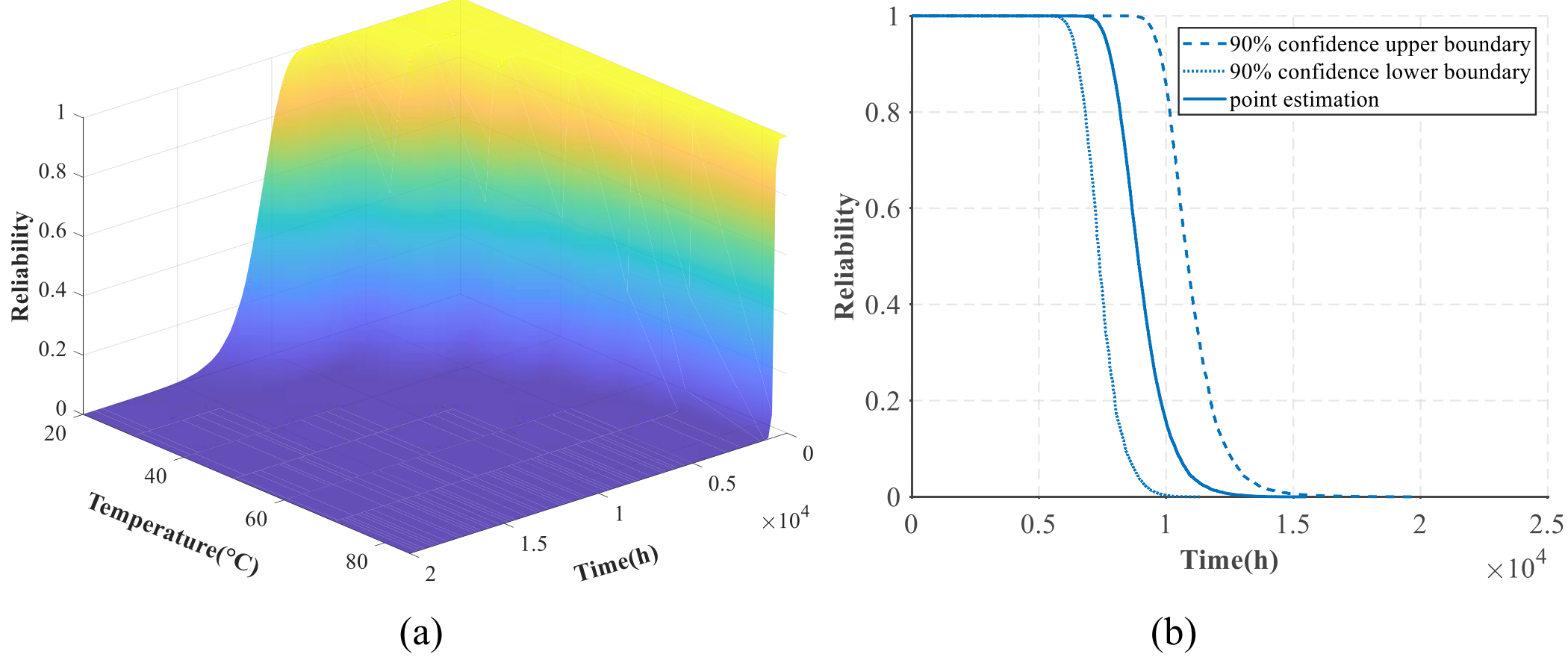

Fig. 7 Reliability of tuner under (a) different temperature and time (b) normal stress level (20°C) with 90% confidence interval.

## 5.3 Discussions

In general, the better the model fits the degradation data, the greater the potential for the model to reflect the degradation law. Besides, it should be recalled that the ultimate goal of ADT modeling is to extrapolate the lifetime and reliability of the product under the normal use condition, so it is crucial to ensure the accuracy of the degradation prediction for unseen stress levels. To this end, we will discuss the fitting and extrapolation capabilities of the proposed model $M_0$ compared to other existing ADT models mentioned in Section 4.3.2.

### 5.3.1 Comparison of model fitting

In this subsection, we first calculate the $l_{\max}$ and the AIC to compare the effectiveness of different ADT models in fitting the degradation data. Table 6 gives the estimations of unknown parameters for the four ADT models. From the $l_{\max}$ and AIC values in Table 6, $M_0$ is the most suitable model compared to other ADT models.

Table 6 Parameter estimations for the four ADT models in the tuner case.

| Model | $\mu_a$ | $\sigma_a$ | $\alpha_1$ | $\beta$ | $\sigma$ | $H$ | $l_{\max}$ | AIC |
|---|---|---|---|---|---|---|---|---|
| $M_0$ | 1.073e-6 | 1.894e-7 | 4.667 | 1.691 | 0.073 | 2.497e-8 | **579.444** | **-1146.888** |
| $M_1$ | 1.349e-6 | / | 4.518 | 1.674 | 0.056 | 0.1113 | 551.784 | -1093.568 |
| $M_2$ | 1.072e-6 | 2.113e-8 | 4.473 | 1.723 | 0.023 | / | 479.348 | -948.696 |
| $M_3$ | 1.601e-6 | / | 4.280 | 1.684 | 0.023 | / | 479.483 | -950.966 |

In addition to evaluating the model's goodness of fit, we are more concerned with the capacity of the model to predict actual degradation processes, which involves deterministic degradation trend predictions and uncertainty quantification under all stress levels. Credible reliability assessments can only be made if the model performs well in both aspects. In order to quantitatively compare the performance of different ADT models in degradation prediction, we set up several indices formulated by Eq. (34) to assess the precision of predicted deterministic degradation trend and uncertainty quantification [40]. Specifically, for each ADT model, we generate 1000 degradation paths according to the estimations in Table 6 and calculate the predicted deterministic degradation trend $x_{lj}^{pre}$, predicted upper boundary $x_{lj}^{pre-U}$ and predicted lower boundary $x_{lj}^{pre-L}$ at $t_{lij}$ under the $l^{\text{th}}$ stress level. Then, the three comparison indices can be calculated by Eq. (34), where $\underset{1\le i\le 1000}{\text{mean}}\left\{x_{lij}^{sim}\right\}$, $\underset{1\le i\le 1000}{\text{quantile}}\left\{x_{lij}^{sim}\right\}_{5\%}^{U}$ and $\underset{1\le i\le 1000}{\text{quantile}}\left\{x_{lij}^{sim}\right\}_{5\%}^{L}$ denote the mean value, upper 5% quantile and lower 5% quantile for the 1000 simulated degradation paths, respectively. The closer $\overline{ER_l}$ is to 0, the more accurately the model predicts the deterministic degradation trend under the $l^{\text{th}}$ stress level. The closer $\overline{ER_l}^U$ or $\overline{ER_l}^L$ is to 0, the better the predicted boundary can describe the uncertainty of the real data under the $l^{\text{th}}$ stress level. Moreover, $\overline{ER}$, $\overline{ER}^U$, and $\overline{ER}^L$ reflect the predicted performance of an ADT model under all stress levels.

$$
\begin{aligned}
&\overline{ER} = \frac{1}{k}\sum_{i=1}^{k}\overline{ER_l}\,,\ \overline{ER_l} = \frac{1}{n_l}\sum_{j=1}^{m_{li}} ER_{lj}\,,\\
&ER_{lj} = \frac{\left| x_{lj}^{pre} - \bar{x}_{lj} \right|}{\bar{x}_{lj}}\,,\ \bar{x}_{lj} = \frac{1}{n_l}\sum_{j=1}^{m_{li}} x_{lij}\,,\ x_{lj}^{pre} = \operatorname*{mean}_{1\le i\le 1000}\left\{ x_{lij}^{sim} \right\};\\
&\overline{ER}^{U} = \frac{1}{k}\sum_{i=1}^{k}\overline{ER_l}^{U}\,,\ \overline{ER_l}^{U} = \frac{1}{n_l}\sum_{j=1}^{m_{li}} ER_{lj}^{U}\,,\\
&ER_{lj}^{U} = \frac{\left| x_{lj}^{pre-U} - x_{lj}^{U} \right|}{x_{lj}^{U}}\,,\ x_{lj}^{U} = \max_{1\le i\le n_l}\left\{ x_{lij} \right\},\ x_{lj}^{pre-U} = \operatorname*{quantile}_{1\le i\le 1000}\left\{ x_{lij}^{sim} \right\}_{5\%}^{U};\\
&\overline{ER}^{L} = \frac{1}{k}\sum_{i=1}^{k}\overline{ER_l}^{L}\,,\ \overline{ER_l}^{L} = \frac{1}{n_l}\sum_{j=1}^{m_{li}} ER_{lj}^{L}\,,\\
&ER_{lj}^{L} = \frac{\left| x_{lj}^{pre-L} - x_{lj}^{L} \right|}{x_{lj}^{L}}\,,\ x_{lj}^{L} = \min_{1\le i\le n_l}\left\{ x_{lij} \right\},\ x_{lj}^{pre-L} = \operatorname*{quantile}_{1\le i\le 1000}\left\{ x_{lij}^{sim} \right\}_{5\%}^{L}.
\end{aligned}
\tag{34}
$$

Table 7 gives the quantitative indices for the four ADT models under all stress levels. The visualization results of the deterministic degradation trend and the degradation boundary are shown in Fig. 8 and Fig. 9.

Table 7 Quantitative indices of the tuner case under all stress levels.

| Model | $\overline{ER}$ | $\overline{ER}^{U}$ | $\overline{ER}^{L}$ |
|---|---|---|---|
| $M_0$ | **0.1335** | **0.9389** | **3.7036** |
| $M_1$ | 0.1490 | 1.1209 | 3.8215 |
| $M_2$ | 0.1429 | 2.3264 | 12.3555 |
| $M_3$ | 0.1849 | 2.4501 | 11.3810 |

*Note*: The bolded results depict the best ones.

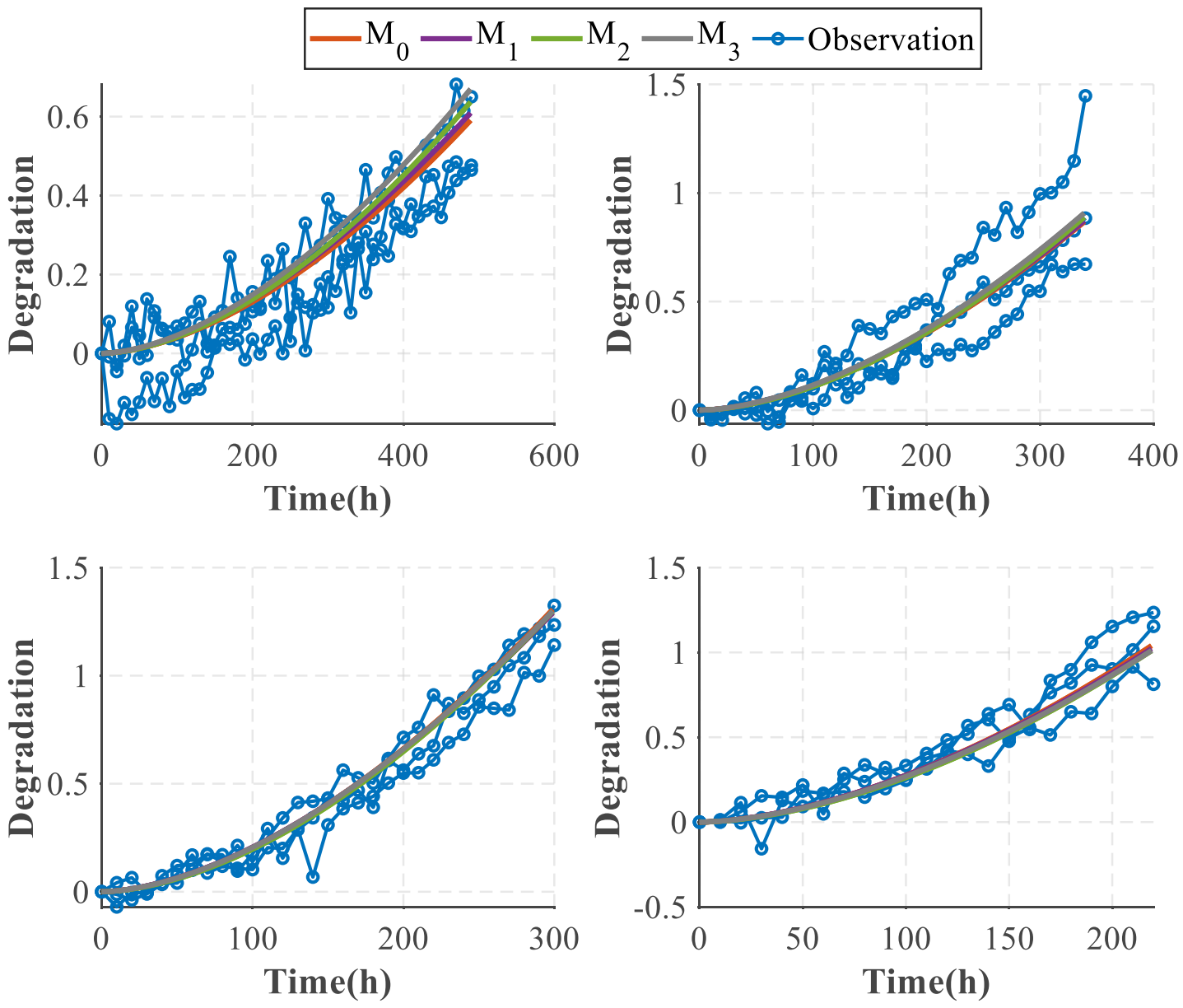

Fig. 8 Prediction for the deterministic degradation trend under all stress levels.

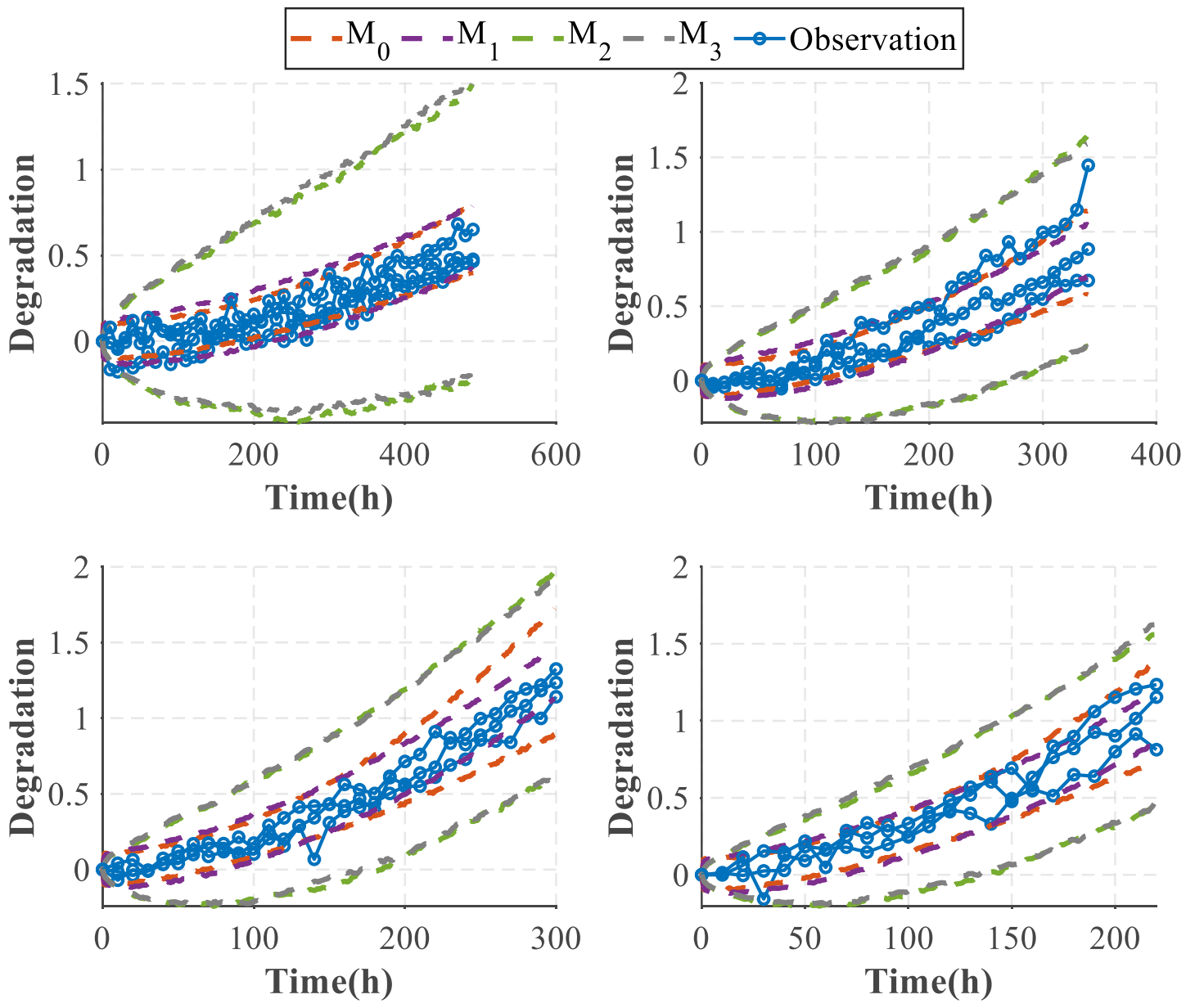

Fig. 9 Prediction for the degradation boundary at 90% CL under all stress levels.

According to the above comparative tables and figures, compared with the other models, model $M_0$ is superior not only in predicting the deterministic degradation trend, but also in uncertainty quantification, which demonstrates the superiority of the proposed model. Additionally, the uncertainty quantification of $M_0$ and $M_1$ is considerably superior to that of models $M_2$ and $M_3$, demonstrating the necessity of considering memory effects in the degradation process.

5.3.2 Comparison of model extrapolation

In this subsection, we develop a cross-validation scheme [41] shown in Table 8 to explore the extrapolation capability of the four ADT models. The results of quantitative indices for the cross-validation 1 and 2 are listed in Table 9 and Table 10, respectively. The visualization results are shown in Fig. 10 and Fig. 11, respectively.

Table 8 Plans of cross-validation

| Projects | Training data | Testing data |
| --- | --- | --- |
| Cross-validation 1 | The data under all stress levels except the lowest stress level | The data under the lowest stress level (55°C) |
| Cross-validation 2 | The data under all stress levels except the highest stress level | The data under the highest stress level (85°C) |

Table 9 Quantitative indices of the tuner case for the cross-validation 1.

| Model | $\overline{ER}_1$ | $\overline{ER}_1^U$ | $\overline{ER}_1^L$ |
| --- | --- | --- | --- |
| $M_0$ | **0.2937** | **0.7971** | 3.3203 |
| $M_1$ | 0.4637 | 1.2940 | **1.3456** |
| $M_2$ | 0.4973 | 3.2261 | 27.1602 |

| $M_3$ | 0.3621 | 3.2290 | 28.5961 |
|---|---|---|---|

*Note*: The bolded results depict the best ones.

Table 10 Quantitative indices of the tuner case for the cross-validation 2.

| Model | $\overline{ER}_4$ | $\overline{ER}_4^U$ | $\overline{ER}_4^L$ |
|---|---|---|---|
| $M_0$ | **0.1249** | **0.4198** | **9.8124** |
| $M_1$ | 0.2167 | 0.4468 | 9.9591 |
| $M_2$ | 0.1617 | 0.8326 | 11.8889 |
| $M_3$ | 0.2153 | 0.7809 | 12.5949 |

*Note*: The bolded results depict the best ones.

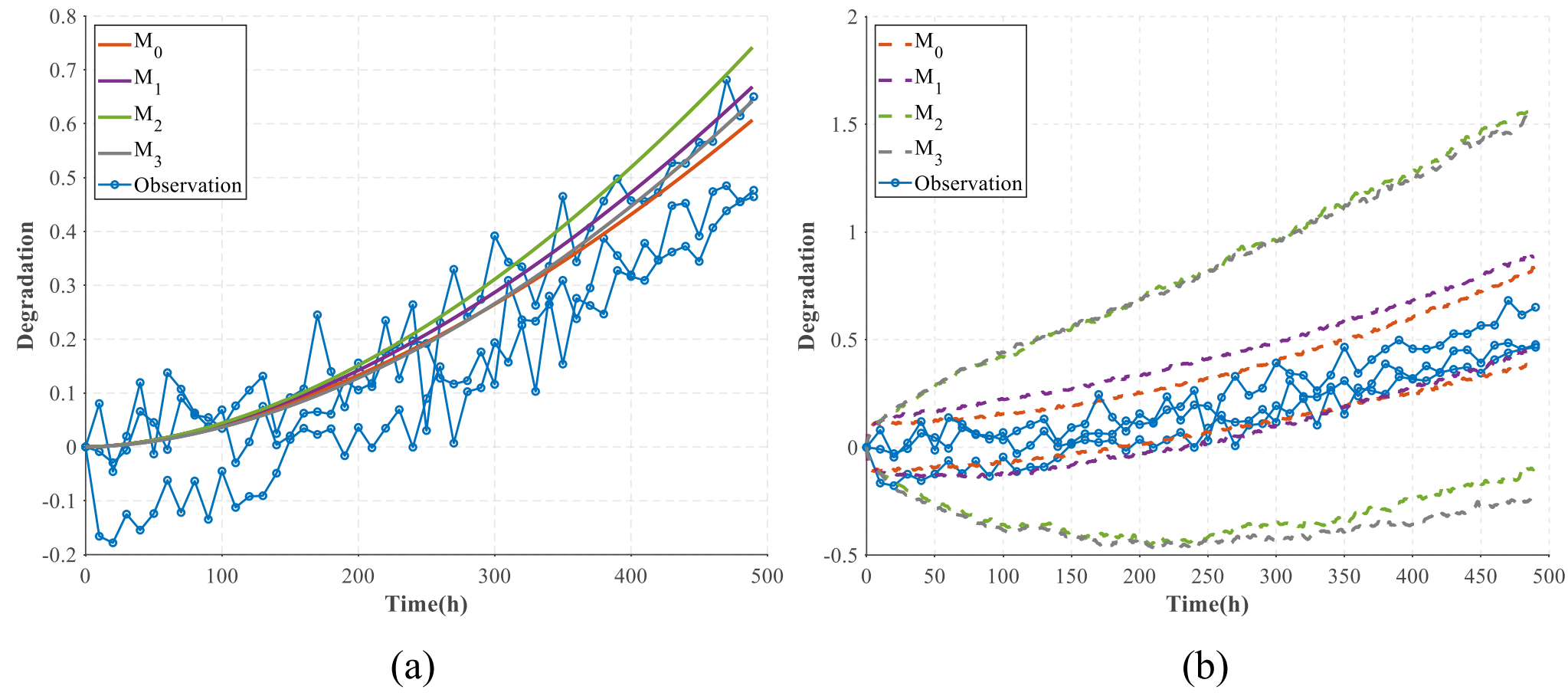

Fig. 10 Prediction for the cross-validation 1: (a) deterministic degradation trend (b) degradation boundary at 90% CL.

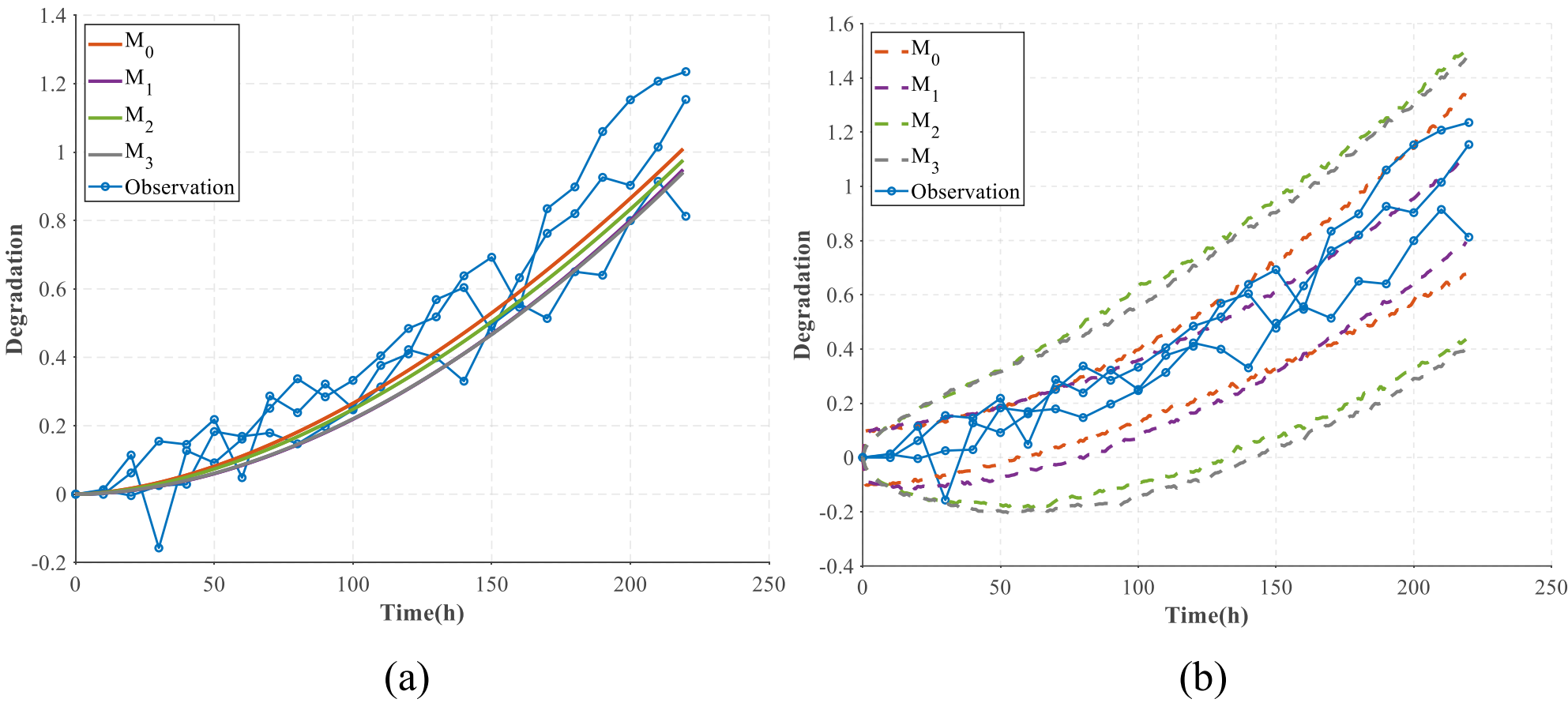

Fig. 11 Prediction for the cross-validation 2: (a) deterministic degradation trend (b) degradation boundary at 90% CL.

From the above comparative tables and figures, the following results can be obtained:

- According to Table 9 and Table 10, for both cross-validations, the predicted deterministic degradation trends based on $M_0$ consistently outperform others significantly. In terms of uncertainty quantification, the upper and lower boundary predictions based on $M_0$ are both the best for cross-validation 2, while the lower boundary predictions based on $M_0$ are worse than those of $M_1$ for cross-validation 1. Overall, model $M_0$ has a superior capacity for uncertainty quantification.

- From Fig. 10 and Fig. 11, for both cross-validations, $M_0$ best describes the deterministic degradation trend and envelopes the real data, proving the superiority of the model $M_0$.
- By analyzing the predictions of these four ADT models for two cross-validations, we can deduce that $M_0$ is more suitable for lifetime and reliability assessment of this tuner than other ADT models.

# 6 Conclusion

This paper seeks to tackle the challenges of reliability modeling and statistical analysis of ADT with memory effects and UtUV. The following conclusions are drawn from this paper:

- This paper proposes a comprehensive degradation model considering the influence of external stresses, memory effects and UtUV. The memory effects are quantified by the Hurst exponent in the FBM and the UtUV is quantified in the acceleration model. Besides, a statistical analysis method based on the EM algorithm is devised to ensure the maximization of the overall likelihood function.
- The results of the simulation case illustrate that the memory effect estimation of the proposed statistical analysis method is more accurate than the two-step MLE method, especially when the number of measurements is small. Moreover, ignoring UtUV will lead to highly biased memory effect estimation and reliability evaluations.
- A tuner case is utilized to demonstrate the efficacy of the proposed methodology. Findings show that the proposed ADT model gives superior predictions in both deterministic degradation trends and uncertainty quantification compared to the existing ADT models, which is more suitable for reliability assessment.

Beyond the work of this study, there exists additional issues that warrant investigation in future research. For example, we only focus on modeling constant-stress ADT data in this work. Constructing an ADT model considering memory effects for the step-stress ADT data is another interesting topic.

# Declarations of interest

None.

# Acknowledgements

This work was supported by the National Natural Science Foundation of China [grant number 51775020], the Science Challenge Project [grant number. TZ2018007], the National Natural Science Foundation of China [grant numbers 62073009].

# Appendix

In this section, we briefly introduce the two-step MLE method [23] for parameter estimations of ADT models with UtUV. The variables used are consistent with those in Section 3.3. In the first step, the parameters associated with the degradation process model are estimated, denoted as $\boldsymbol{\theta}_1 = [\beta, \sigma, H]$. Subsequently, the remaining parameters relevant to the acceleration model are determined in the second step, denoted as $\boldsymbol{\theta}_2 = [\mu_a, \sigma_a, \alpha_1]$.

In the first step, based on the property of FBM, for the $i^{th}$ item tested at $l^{th}$ stress level, we have

$$\boldsymbol{x}_{li} \sim N\left(e_{li}\boldsymbol{\tau}_{li}, \sigma^2\boldsymbol{\Sigma}_{li}\right), \tag{35}$$

where

$$e_{li} = a_{li}\exp\left(\alpha_1 s_l^*\right). \tag{36}$$

Based on Eq. (35), the log-likelihood function of the ADT data can be derived as:

$$\ln L_0\left(\boldsymbol{\theta}|\boldsymbol{x}\right) = -\frac{1}{2}\sum_{l=1}^{k}\sum_{i=1}^{n_l}\left\{m_{li}\ln(2\pi) + \ln\left|\sigma^2\boldsymbol{\Sigma}_{li}\right| + \frac{\left(\boldsymbol{x}_{li} - e_{li}\boldsymbol{\tau}_{li}\right)^T\boldsymbol{\Sigma}_{li}^{-1}\left(\boldsymbol{x}_{li} - e_{li}\boldsymbol{\tau}_{li}\right)}{\sigma^2}\right\} \tag{37}$$

Then, the first partial derivatives of $\ln L_0$ with respect to $e_{li}$ and $\sigma^2$ can be derived, respectively. By equating each derivative to zero, we can obtain

$$\hat{e}_{li} = \frac{\boldsymbol{x}_{li}^T\boldsymbol{\Sigma}_{li}^{-1}\boldsymbol{\tau}_{li}}{\boldsymbol{\tau}_{li}^T\boldsymbol{\Sigma}_{li}^{-1}\boldsymbol{\tau}_{li}}, \tag{38}$$

$$\hat{\sigma}^2 = \frac{\sum_{l=1}^{k}\sum_{i=1}^{n_l}\left(\boldsymbol{x}_{li} - \hat{e}_{li}\boldsymbol{\tau}_{li}\right)^T\boldsymbol{\Sigma}_{li}^{-1}\left(\boldsymbol{x}_{li} - \hat{e}_{li}\boldsymbol{\tau}_{li}\right)}{\sum_{l=1}^{k}\sum_{i=1}^{n_l}m_{li}}. \tag{39}$$

Upon substituting Eqs (38) and (39) into Eq. (37), the log-likelihood function depends solely on $\beta$ and $H$. Thus, $\hat{\beta}$ and $\hat{H}$ can be obtained by a two-dimensional search for the maximum value of Eq. (37). Then, $\hat{e}_{li}$ and $\hat{\sigma}^2$ can calculated by substituting $\hat{\beta}$ and $\hat{H}$ into Eqs (38) and (39).

In the second step, according to Eq. (11), we can get

$$e_{li} \sim N\left(\mu_a\exp\left(\alpha_1 s_l^*\right), \sigma_a^2\exp\left(2\alpha_1 s_l^*\right)\right) \tag{40}$$

Based on Eq. (40), the log-likelihood function for $e_{li}$ can be derived as:

$$\ln L_0\left(\boldsymbol{\theta}|\boldsymbol{x}\right) = -\frac{1}{2}\sum_{l=1}^{k}\sum_{i=1}^{n_l}\left\{m_{li}\ln(2\pi) + \ln\sigma_a^2 + 2\alpha_1 s_l^* + \frac{\left[e_{li} - \mu_a\exp\left(\alpha_1 s_l^*\right)\right]^2}{\sigma_a^2\exp\left(2\alpha_1 s_l^*\right)}\right\} \tag{41}$$

Specifically, $\hat{\mu}_a$ and $\hat{\sigma}_a^2$ can be computed as:

$$\hat{\mu}_a = \frac{1}{\sum_{l=1}^{k}n_l}\sum_{l=1}^{k}\sum_{i=1}^{n_l}\frac{\hat{e}_{li}}{\exp\left(\alpha_1 s_l^*\right)}, \tag{42}$$

$$\hat{\sigma}_a^2 = \frac{1}{\sum_{l=1}^{k}n_l}\sum_{l=1}^{k}\sum_{i=1}^{n_l}\left(\frac{\hat{e}_{li}}{\exp\left(\alpha_1 s_l^*\right)} - \hat{\mu}_a\right)^2 \tag{43}$$

By substituting Eqs (42) and (43) into Eq. (41), the log-likelihood function depends solely on $\alpha_1$. Therefore, $\hat{\alpha}_1$ can be obtained by a one-dimensional search for the maximum value of Eq. (41). Then, $\hat{\mu}_a$ and $\hat{\sigma}_a^2$ can calculated by substituting $\hat{\alpha}_1$ into Eqs (42) and (43).